\documentclass[12pt]{article}
\usepackage{a4wide}
\usepackage{amssymb}
\usepackage{amsmath}
\usepackage{graphicx}
\usepackage{mathrsfs}
\usepackage{slashed}
\usepackage{verbatim}
\usepackage{mathtools}
\usepackage{hyperref}

\def\XXint#1#2#3{{\setbox0=\hbox{$#1{#2#3}{\int}$}
     \vcenter{\hbox{$#2#3$}}\kern-.52\wd0}}

\newcommand{\be}{\begin{equation}}\newcommand{\ee}{\end{equation}}
\newcommand{\bea}{\begin{eqnarray}} \newcommand{\eea}{\end{eqnarray}}

\def\makeatletter{\catcode`\@=11}
\makeatletter
\def\mathbox#1{\hbox{$\m@th#1$}}%
\def\math@ccstyles#1#2#3#4#5#6#7{{\leavevmode
      \setbox0\mathbox{#6#7}%
      \setbox2\mathbox{#4#5}%
      \dimen@ #3%
      \baselineskip\z@\lineskiplimit#1\lineskip\z@
      \vbox{\ialign{##\crcr
             \hfil \kern #2\box2 \hfil\crcr
             \noalign{\kern\dimen@}%
             \hfil\box0\hfil\crcr}}}}
\def\mathaccstyles{\math@ccstyles\maxdimen}
\def\maththroughstyles{\math@ccstyles{-\maxdimen}}
\def\unity%
 {\maththroughstyles{.45\ht0}\z@\displaystyle {\mathchar"006C}\displaystyle 1}

\makeatletter \@addtoreset{equation}{section} \makeatother

\begin{document}

\setcounter{table}{0}

\begin{flushright}\footnotesize

\texttt{ICCUB-17-001}
\vspace{0.6cm}
\end{flushright}

\mbox{}
\vspace{0truecm}
\linespread{1.1}

\centerline{\LARGE \bf Large $N$ correlation functions  }

\vspace{.5cm}

 \centerline{\LARGE \bf in $\mathcal{N}=2$ superconformal  quivers}

\vspace{1.5truecm}

\centerline{
    {\large \bf Alessandro Pini${}^{a}$} \footnote{pinialessandro@uniovi.es}
    {\large \bf Diego Rodriguez-Gomez${}^{a}$} \footnote{d.rodriguez.gomez@uniovi.es}
    {\bf and}
    {\large \bf Jorge G. Russo ${}^{b,c}$} \footnote{jorge.russo@icrea.cat}}

\vspace{1cm}
\centerline{{\it ${}^a$ Department of Physics, Universidad de Oviedo}} \centerline{{\it Avda.~Calvo Sotelo 18, 33007  Oviedo, Spain}}
\medskip
\centerline{{\it ${}^b$ Instituci\'o Catalana de Recerca i Estudis Avan\c{c}ats (ICREA)}} \centerline{{\it Pg.Lluis Compayns, 23, 08010 Barcelona, Spain}}
\medskip
\centerline{{\it ${}^b$ Departament de F\' \i sica Cu\' antica i Astrof\'\i sica and Institut de Ci\`encies del Cosmos}} \centerline{{\it Universitat de Barcelona}}\centerline{{\it Mart\'i Franqu\`es, 1, 08028
Barcelona, Spain }}
\vspace{1cm}

\centerline{\bf ABSTRACT}
\medskip

Using supersymmetric localization, we
consider four-dimensional $\mathcal{N}=2$ superconformal quiver gauge theories  obtained from $\mathbb{Z}_n$ orbifolds of $\mathcal{N}=4$ Super Yang-Mills theory in the large $N$ limit at weak coupling.
In particular, we show that: 1) The partition function for arbitrary couplings can be constructed in terms of universal building blocks.
2) It can be computed in perturbation series, which
converges uniformly for $|\lambda_I|<\pi^2$, where
$\lambda_I$ are the 't Hooft coupling of the gauge groups.
3) The perturbation series for two-point functions can be explicitly computed to arbitrary orders. 
There is no universal effective coupling
by which one can express them in terms of correlators of the $\mathcal{N}=4$ theory.
4) One can define  twisted and untwisted sector operators. At the perturbative orbifold point, when all the couplings are the same, the correlators of untwisted sector operators coincide with those of $\mathcal{N}=4$ Super Yang-Mills theory. In the twisted sector, we find remarkable cancellations of a certain number of planar loops, determined by  the conformal dimension
of the operator.

\noindent

\newpage

\tableofcontents

\section{Introduction}

Among the most remarkable discoveries of the last decades, the AdS/CFT duality occupies a prominent place and has led to multiple applications and research lines.
 The best known example of the duality is $\mathcal{N}=4$ Super Yang-Mills (SYM) theory, which provides a holographic description of superstring theory on the $AdS_5\times S^5$ space.
This case has withstood numerous detailed tests, to great extent, by exploiting integrability of the supersymmetric gauge theory in the planar limit and supersymmetric localization \cite{Pestun:2007rz}.
While there are other examples of superconformal field theories with exact gravity duals that have been thoroughly studied -- such as ABJM theory or $\beta $-deformed  $\mathcal{N}=1$ SYM -- 
comparatively much less is understood to what concerns the detailed aspects of how AdS/CFT duality works in more general $\mathcal{N}=2$ four-dimensional superconformal gauge theories. 

A well-known example is $\mathcal{N}=2$ superconformal quantum chromodynamics (SCQCD), with gauge group $SU(N)$ coupled to $2N$ flavor hypermultiplets. However, this case does not appear to have a simple string theory dual. Indeed, while there has been some suggested dual backgrounds in the literature (see \textit{e.g.} \cite{Gaiotto:2009gz,Gadde:2009dj,Gadde:2010zi}), none of these is of the form $AdS_5\times M_5$ where one can perform calculations in a controlled classical gravity approximation. Far from this, the string theory dual seems to be in the deep quantum regime. In particular, one sign of this is the fact that the string tension appears to be proportional to the logarithm of the 't Hooft coupling \cite{Passerini:2011fe}. 

There is, however, a whole family of $\mathcal{N}=2$ four-dimensional superconformal
field theories with simple gravity duals which contain, in a certain limit, $\mathcal{N}=2$ superconformal QCD. These are the so-called necklace quivers, with $n$ nodes and $SU(N)$ gauge groups, joined by bifundamental hypermultiplets. These theories can be engineered by $N$ D3 branes probing an $A_{n-1}$ singularity; which, in the suitable near-brane limit, are dual to superstring theory on the orbifold on $AdS_5\times S^5/\mathbb{Z}_n $. The simplest example is $n=2$, for which we have an $SU(N)\times SU(N)$ gauge theory
with two bifundamental hypermultiplets.

It is then natural to wonder to what extent the quiver theory --with a regular gravity dual-- shares common properties with $\mathcal{N}=2$ SCQCD , with the hope that this may help to better understand main differences with respect to a gauge theory dual to a regular string background (studies along these lines have been initiated in \cite{Gadde:2009dj,Gadde:2010zi}). 

In superconformal field theories, very important observables are the  correlation functions. A particularly interesting set of operators are chiral primary operators (CPO's). It turns out that correlators of CPO's exhibit a very interesting structure. This has been explored in the case of $\mathcal{N}=2$ SCQCD  in a beautiful series of papers \cite{Papadodimas:2009eu,Baggio:2014sna,Baggio:2014ioa,Baggio:2015vxa}. 
In this paper we will compute, for the first time, general two-point correlation functions of chiral primary operators (CPO's) in quiver gauge  theories. Our main tool will be supersymmetric localization. Very recently, it was argued  \cite{Gerchkovitz:2016gxx} that correlation functions on $\mathbb{R}^4$ of CPO's in $\mathcal{N}=2$ superconformal theories admitting a Lagrangian description can be computed as correlation functions of the associated operators in the  matrix model that describes the theory on $\mathbb{S}^4$. The subtlety stands in that, upon mapping the theory to the $\mathbb{S}^4$, due to the conformal anomaly, a non-trivial mixing structure among operators of different dimension is induced. The insight of \cite{Gerchkovitz:2016gxx} is that such mixture can be disentangled by a Gram-Schmidt procedure. Using this prescription, in
\cite{Rodriguez-Gomez:2016ijh,Rodriguez-Gomez:2016cem,Baggio:2016skg} correlation functions for CPO's in $\mathcal{N}=4$ and $\mathcal{N}=2$ SCQCD  were computed. Moreover, it turns out that these operator mixtures have a very interesting structure, as shown in \cite{Rodriguez-Gomez:2016cem,Baggio:2016skg}, whose holographic interpretation is still to be understood.

In this paper we will compute, using supersymmetric localization as well as the Gram-Schmidt procedure proposed in \cite{Gerchkovitz:2016gxx}, correlation functions of CPO's in the $A_{n-1}$ quivers in the large $N$ theory at weak 't Hooft coupling. 
A given $n$-loop order is, in general, a power $\lambda^n/(2\pi)^{2n}$
multiplied by a combination of products of Riemann $\zeta $ coefficients and rational numbers.
In particular, there is an  infinite series of terms which have linear dependence with $\zeta(2k-1)$ that can be safely isolated and studied, as they
are independent irrational numbers that can be distinguished from the rest of the terms. As we show, this method can be applied to the partition function itself. Interestingly, a ``modular" structure appears, as the partition function ``factorizes" into contributions from the different quiver nodes. Of course, when the couplings are equal, we recover the orbifold equivalence first found in \cite{Azeyanagi:2013fla}.
We should stress that this structure emerges after the functional integration, that is, in the final result.

The structure of this paper is as follows. In section \ref{sec1} we give a lightning review of the $A_{n-1}$ necklace theories of interest and compute, in subsection \ref{sectpartitionfunction}, their partition function, in particular, exhibiting its factorization property. In the rest of the section, we discuss its
convergence properties and its implications for the holographic correspondence.
In section \ref{sec2} we compute the two-point  correlation functions on $\mathbb{S}^4$ for CPO's. 
In subsection \ref{secQCD} we compute the correlation functions  for CPO's in $\mathcal{N}=2$ superconformal QCD including all  corrections which have linear $\zeta(2k-1)$ dependence. In section \ref{secquiver} we turn to the computation of the correlators on $\mathbb{S}^4$ in the quiver gauge theories. 
In section 3.3 we work out  the $A_1$ case in detail and compute the 
 correlators on $\mathbb{R}^4$. 
 Section \ref{conclusions} contains a summary of the results and concluding remarks.
As  computations  are rather long, we collect their details in a number of appendices. In appendix \ref{Notation} we summarize the notation. In appendix \ref{correlatorsN=4} we collect the expressions for large $N$ correlation functions in $\mathcal{N}=4$ SYM computed in \cite{Rodriguez-Gomez:2016ijh}. 
In appendix \ref{appendixZA1} we describe  the computation of the partition function. In appendix \ref{correlatorsN=2} we give the details of  the computation of correlators in $\mathcal{N}=2$ superconformal QCD including all corrections linear in $\zeta$ as well as the first non-linear correction proportional to $\zeta(3)^2$. Finally, in appendix \ref{correlatorsquiver} we provide further details on the computation of the correlation functions in the quiver gauge theories.

\section{Necklace $\mathcal{N}=2$ superconformal quiver theories}\label{sec1}

We will be interested on superconformal gauge theories with four-dimensional $\mathcal{N}=2$ supersymmetry which arise as  $\mathbb{Z}_n$ orbifolds of $\mathcal{N}=4$ SYM \cite{Douglas:1996sw}. These theories can be engineered in string theory on the worldvolume of $N$ D3 branes probing a $\mathbb{C}^2/\mathbb{Z}_n$ singularity. In fact, in the suitable limit, they admit  a weakly curved gravity dual
in terms of the $AdS_5\times S^5/\mathbb{Z}_n$ geometry.

\begin{figure}[h!]
\centering
\includegraphics[scale=.6]{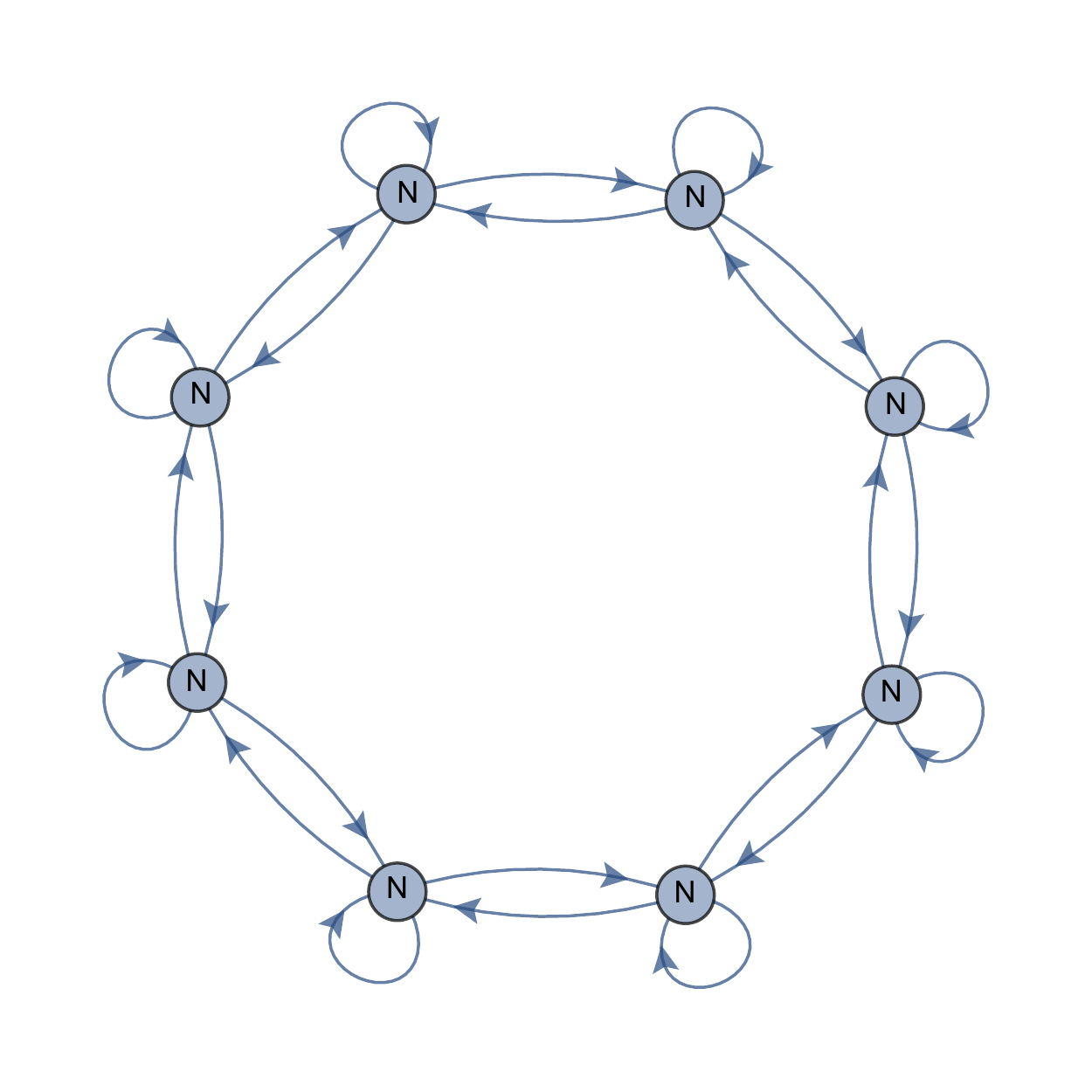}
\caption{$A_{n-1}$ necklace quiver gauge  theories.}
\label{quiver}
\end{figure}

These quiver theories can be represented pictorially as
in fig. \ref{quiver} (we show the  $n=8$ case, as example).

Each node stands for a $SU(N)$ vector multiplet\footnote{One may consider the general case of a gauge group $\prod_i SU(N_i)$. However, superconformal invariance requires all $N_i=N$.}, while each arrow connecting nodes $I$ and $J$ corresponds a chiral multiplet in the bifundamental of $SU(N)_I\times SU(N)_J$ --when $I=J$ this corresponds to an adjoint. Denoting each chiral multiplet between node $I$ and $J$ as $X_{I,J}$, the superpotential reads

\begin{equation}
W=\sum_{I=1}^n {\rm Tr} X_{I,I}X_{I,I+1}X_{I+1,I}-{\rm Tr} X_{I-1,I-1}X_{I-1,I}X_{I,I-1}\, .
\end{equation}
where the node $n+1$ is identified with the node 1.

This theory is conformal for arbitrary couplings $g_I$. 
The perturbative orbifold corresponds to the case $g_I=g$ for all nodes. In the dual interpretation,  $g_I^{-2}-g_{I+1}^{-2}\sim \int_{\Sigma_I} B_2$, and $\Sigma_I$ represents  the appropriate collapsed 2-cycle in the geometry. \footnote{As it is well known, the perturbative orbifold contains a non-zero $B_2$. For instance, in the $\mathbb{Z}_2$ case  \cite{Aspinwall:1994ev}, one has $\int_{\Sigma} B_2 =\frac{1}{2}$ and, strictly speaking, one should write $g_1^{-2}-g_{2}^{-2}\sim \int_{\Sigma} B_2-\frac{1}{2}$.}

Note that each node looks like a copy of $\mathcal{N}=2$ superconformal QCD with coupling $g_I$, as the two neighbouring nodes supply the $2N$ flavors needed to make the $SU(N)$ gauge group conformal. Thus, it then follows that in the limit in which all but one coupling --say $g_1$-- go to zero we recover $\mathcal{N}=2$ superconformal QCD. It is in this sense that the quiver theory interpolates between a theory --the $A_{n-1}$ quiver-- with a regular gravity dual and $\mathcal{N}=2$ superconformal QCD.

\subsection{The partition function for necklace quiver theories}\label{sectpartitionfunction}

The partition function (or more precisely, its logarithm) for $\mathcal{N}=2$ superconformal field theories has been shown to compute the K\" ahler  potential $K$ on the conformal manifold \cite{Gerchkovitz:2014gta,Gomis:2014woa}. As such, it is subject only to K\" ahler  transformations $K(\tau,\overline{\tau})\rightarrow  K(\tau,\overline{\tau}) + \mathcal{F}(\tau)+\overline{\mathcal{F}}(\overline{\tau})$. Hence, its (non-analytic) dependence on the 't Hooft coupling $\lambda\sim ({\rm Im}\tau)^{-1}$ is a physically unambiguous quantity.

The exact partition function for the $A_{n-1}$ necklace quiver gauge theories of interest can be easily constructed using the general results of \cite{Pestun:2007rz}. It reads

\begin{equation}
\label{pathintegralZAn-1}
Z_{A_{n-1}}= \int \prod_{I=1}^n \left( d^{N-1}a^I_i \,\Delta_I \ e^{- \frac{8\pi^2}{g_I^2}\sum_{i=1}^N (a^I_i)^2} \right)\, Z_{\rm 1-loop}\,Z_{\rm inst}
\, ,\qquad \Delta_I= \prod_{i<j}(a^I_i-a^I_j)^2\, ,
\end{equation}
where the variables $a_i^I$  stand for the eigenvalues of the adjoint scalar in the vector multiplet $X_{I,I}$. $Z_{\rm inst}$ stands for the instanton contribution and $Z_{\rm 1-loop}$ is the one-loop contribution given by

\begin{equation}
Z_{\rm 1-loop}=\frac{\prod_{I} \prod_{i< j} H^2(a^I_i-a^I_j)}{\prod_{I} \prod_{i,j} H(a^I_i-a^{I+1}_j)}\, ;
\end{equation}
where the node $n+1$ is identified with the  node 1. Note that, as the gauge groups are  $SU(N)$,  one has $\sum_i a_i^I=0$. 

The function $H$ can be written in terms of Barnes G-functions. For the purpose of this paper, we will define it
in terms of the Taylor expansion
\begin{equation}
\label{expansionH}
\ln H(x)=-\sum_{n=2}^{\infty}\frac{(-1)^n}{n}\zeta(2n-1)x^{2n}\, ,
\end{equation}
which converges for $|x|<1$.

In the following we will be interested on the large $N$ limit of the quiver theories, with $\lambda_I \equiv N g_I^2$ fixed. 
We will assume, as usual, that instantons  are suppressed, as their contributions are multiplied by an exponentially small factor $e^{-\frac{8\pi^2 |k| N}{\lambda_I}}$, with integer $k$. While there have  been explicit checks in some $\mathcal{N}=2$
theories (see e.g. \cite{Passerini:2011fe,Russo:2013kea}), for the necklace quiver theories the suppression of instantons at large $N$ remains a plausible assumption that is yet to be studied (see \cite{Azeyanagi:2013fla} however for a more detailed account).
 Therefore from now on we set  $Z_{\rm inst}=1$.

\subsection{The case of the $A_1$ theory}\label{A1}

For concreteness, let us concentrate on the $A_1$ case, corresponding to $n=2$. This describes an $\mathcal{N}=2$ superconformal gauge theory with gauge group $SU(N)\times SU(N)$ and two bifundamental chiral multiplets. 
Some aspects of this theory were discussed in  \cite{Gadde:2009dj,Gadde:2010zi,Russo:2013kea,Pomoni:2013poa,Mitev:2014yba,Fraser:2015xha,Mitev:2015oty}. Re-naming $a_i^1\rightarrow a_i$ and $a_i^2\rightarrow b_i$, the partition function reads

\begin{equation}
\label{ZA1}
Z_{A_1}=\int d^{N-1}a_i \int d^{N-1}b_i \,\Delta(a)\Delta(b) Z_{\rm 1-loop}\  e^{-\frac{8\pi^2 N}{\lambda_1^2}\sum a_i^2}e^{-\frac{8\pi^2N}{\lambda_2^2}\sum b_i^2}\, ;
\end{equation}
where

\begin{equation}
\Delta(a)= \prod_{i<j}(a_i-a_j)^2\, \qquad \Delta(b)= \prod_{i<j}(b_i-b_j)^2\ ,
\end{equation}
while 

\begin{equation}
Z_{\rm 1-loop}=\frac{\prod_{i< j} H^2(a_i-a_j)\prod_{i<j}H^2(b_i-b_j)}{\prod_{i,j} H^2(a_i-b_j)}\, .
\end{equation}

Let us consider the weak coupling regime $\lambda_I\ll 1$ for all nodes. In this regime the largest contribution to the partition function comes from the region of small $a_i^I$, $b_i^I$. We can then use the Taylor expansion (\ref{expansionH}) of $\ln H(x)$ for small $x$.
This expansion gives rise to the perturbation series.

The $\mathcal{N}=2$ quiver partition function can be fully written in
terms of correlators of the $\mathcal{N}=4$ matrix model. For this, it is convenient to write $Z_{\rm 1-loop}$ as

\begin{equation}
\label{eef}
Z_{\rm 1-loop}=e^{f}\, ;\qquad f=\sum_{i,j} \left(\ln H(a_i-a_j)+\ln H(b_i-b_j)-2\ln H(a_i-b_j)\right)\, .
\end{equation}

Expanding $e^{f}=1+f+...$, the first term
``$1$" gives rise to a contribution which is the product of two $\mathcal{N}=4$ partition function
with couplings $\lambda_1$ and $\lambda_2$.
The remaining terms give rise to the correlators which are shown in detail  in appendix \ref{appendixZA1}. In particular, it is useful to introduce the correlators in the $\mathcal{N}=4$ matrix model\footnote{We shall use the notation $\langle\langle \cdot\rangle\rangle$ to denote correlators in the corresponding $\mathbb{S}^4$ matrix model. In turn, by $\langle\langle \cdot\rangle\rangle_0$ we shall denote correlators in the $\mathcal{N}=4$ matrix model. We shall also use the notation  $\sum a_i^n\equiv X_n$. See appendix \ref{Notation} for a summarized description of the notation.}

\begin{eqnarray}
\label{Cs}
\langle\langle X_n\rangle\rangle_0 = C_n(g_I)&=&\frac{1}{Z_{\mathcal{N}=4}(g_I)}\int d^{N-1}a\ e^{-\frac{8\pi^2}{g_I^2} \sum a_i^2} \Big(\sum_i a_i^n\Big)\,, \\  \langle\langle X_n X_m\rangle\rangle_0=C_{m,n}(g_I)&=&\frac{1}{Z_{\mathcal{N}=4}(g_I)}\int d^{N-1}a\  e^{-\frac{8\pi^2}{g_I^2} \sum a_i^2}\, \Big(\sum_i a_i^m\Big)\,\Big(\sum_i a_i^n\Big)\, ;
\end{eqnarray}
with (see \textit{e.g.} \cite{Drukker:2000rr})

\begin{equation}
\label{ZN4}
Z_{\mathcal{N}=4}=\mathcal{Z}_{\mathcal{N}=4}^{SU(N)}(g) =\sqrt{\frac{8\pi N}{g^2}}\Big(\frac{16\pi^2}{g^2} \Big)^{-\frac{N^2}{2}}\, \,(2\pi)^{\frac{N}{2}}\, G(N+2)\, .
\end{equation}
We can also define the connected correlators in the $\mathcal{N}=4$ SYM 
$$
\langle\langle X_{m_1}\cdots X_{m_n}\rangle\rangle_0^c \equiv \mathscr{C}_{m_1\cdots m_n}\ .
$$

For these correlators, it is useful to introduce a deformed version of the $\mathcal{N}=4$ theory by adding  sources for all single-trace chiral primary operators  \cite{Gerchkovitz:2016gxx}

\begin{equation}
{Z}_{\mathcal{N}=4}(g_I,\,\{h_{i,A}\})=\int d^{N-1}a \,\Delta(a)\,e^{-\frac{8\pi^2}{g_I^2}\sum_ia_i^2+\sum_{A=3}^N h_{I,A} \sum_i  a_i^A}\ .
\end{equation}
The connected correlators $\langle\langle X_{m_1}\cdots X_{m_n}\rangle\rangle_0^c$ can then be computed from the associated free energy $F=- \ln{Z}_{\mathcal{N}=4}$ as

\begin{equation}
\label{mathscrCs}
\langle\langle X_{m_1}\cdots X_{m_n}\rangle\rangle_0^c= \mathscr{C}_{m_1,\cdots,m_n}(g_I)=\frac{\partial}{\partial h_{I,m_1}}\cdots \frac{\partial}{\partial h_{I,m_n}} F(g_I,\,\{h_{I,A}\})\Big|_{\{h_{I,A}=0\}}\ .
\end{equation}
It is straightforward to re-write the $C_{m_1,\cdots,m_n}$'s in terms of the $\mathscr{C}_{m_1,\cdots,m_n}$'s, whose expression has been computed (for instance, in this context and for 2-,3- and 4-points, in \cite{Rodriguez-Gomez:2016ijh}). In particular, in the planar limit, a simple calculation shows that 
$\mathscr{C}_{m_1\cdots m_n}$ scales as

\begin{equation}
\mathscr{C}_{m_1\cdots m_n}\sim \frac{1}{N^{n-2}}\, .
\end{equation}
Taking into account this,  we can compute the leading term in $N$, finding (see appendix \ref{appendixZA1} for details)

\begin{equation}
\label{partiZZ}
\ln Z_{A_1} = \ln Z_{\mathcal{N}=4}(\lambda_1)
+\ln Z_{\mathcal{N}=4}(\lambda_2) + f_I \, ,
\end{equation}
where, modulo an irrelevant additive, $\lambda$-independent constant (see \eqref{ZN4})
\begin{equation}
\ln Z_{\mathcal{N}=4}(\lambda)=\frac{N^2}{2}\ln \lambda
\end{equation}
and
\begin{equation}
\label{fZZ}
 f_I=-\sum_{n=2}^\infty\sum_{k=0}^{2n} \frac{(-1)^{n+k}}{n}\zeta(2n-1) {{2n}\choose{k}} (\mathscr{C}_{2n-k}(\lambda_1)-\mathscr{C}_{2n-k}(\lambda_2))(\mathscr{C}_{k}(\lambda_1)-\mathscr{C}_{k}(\lambda_2))\, .
\end{equation}
Separating the $k$ sum into odd and even, the odd part trivially vanishes since $\mathscr{C}_{2n+1}=0$
(see appendix  appendix \ref{correlatorsN=4}). Thus 

\begin{equation}
 f_I=-\sum_{n=2}^\infty \sum_{k=1}^{n-1} \frac{(-1)^{n}}{n}\zeta(2n-1) {{2n}\choose{2k}} (\mathscr{C}_{2n-2k}(\lambda_1)-\mathscr{C}_{2n-2k}(\lambda_2))(\mathscr{C}_{2k}(\lambda_1)-\mathscr{C}_{2k}(\lambda_2))\, .
\end{equation}
Introducing

\begin{equation}
\label{Fhyper}
\mathfrak{F}(\lambda_1,\lambda_2)=e^{\sum^{\infty}_{n=2}\sum_{k=1}^{n-1} \frac{(-1)^{n}}{n}\zeta(2n-1) {{2n}\choose{2k}} \mathscr{C}_{2n-2k}(\lambda_1)\mathscr{C}_{2k}(\lambda_2)}\, ,
\end{equation}
we have that
\begin{equation}
e^{f_I}=\frac{\mathfrak{F}(\lambda_1, \lambda_2)\mathfrak{F}(\lambda_2,\lambda_1)}{\mathfrak{F}(\lambda_1,\lambda_1)\mathfrak{F}(\lambda_2,\lambda_2)}\,.
\end{equation}
Using the expressions for $\mathscr{C}_{2k}$ given in appendix \ref{correlatorsN=4}, we finally find
\be
\label{buildingblock}
 \ln {\mathfrak{F}}(\lambda_1,\lambda_2)  = 2N^2 \sum_{n=2}^\infty  
 \frac{ (-1)^n (2n-1)!}{(4\pi )^{2n}}  \zeta(2n-1) \sum_{k=1}^{n-1} \frac{\lambda_1^{n-k} \lambda_2^k }{k!(k+1)!(n-k)!(n-k+1)!}\, 
 \ee
Thus

\begin{equation}
Z_{A_1}=Z_{\mathcal{N}=4}(\lambda_1)Z_{\mathcal{N}=4}(\lambda_2)\frac{\mathfrak{F}(\lambda_1,\lambda_2)\mathfrak{F}(\lambda_2,\lambda_1)}{\mathfrak{F}(\lambda_1,\lambda_1)\mathfrak{F}(\lambda_2,\lambda_2)}\,.
\label{Zfactor2}
\end{equation}

\subsection{The general case of $A_n$ quiver gauge theories}

Coming to the general case, let us write the 1-loop contribution as

\begin{equation}
Z_{\rm 1-loop}=\prod_{I} \frac{ \prod_{i< j} H^2(a^I_i-a^I_j)}{\prod_{i,j} H(a^I_i-a^{I+1}_j)}=\prod_Ie^{\sum_{i,j}\ln H(a_i^I-a_j^I)-\ln H(a_i^I-a_j^{I+1})}\, .
\end{equation}
Using the weak coupling expansion  \eqref{expansionH} and  
$$
\sum_i (a_i^I)^n={\rm Tr}X^n_{I,I}\equiv X_I^n \, ,
$$ 
(see appendix \ref{Notation} for the notation), we can write
\begin{equation}
\label{generaln}
Z_{\rm 1-loop}=\prod_{I}e^{-\sum_{n=2}^\infty\sum_{k=0}^{2n}\frac{(-1)^{n+k}}{n}\zeta(2n-1)\binom{2n}{k}
\left( X_{I}^{2n-k}X_{I}^k - X_{I}^{2n-k}X_{I+1}^k \right)}\, .
\end{equation}
We now substitute \eqref{generaln} into the partition function \eqref{pathintegralZAn-1} and expand the exponential, just as in the $n=2$ case. At large $N$, the leading terms come from the one-point functions. This amounts to trading ${\rm Tr}X_{I,I}^a$ for $\mathscr{C}_a(g_I)$, so that in \eqref{generaln} we easily recognize the function $\mathfrak{F}(\lambda_I,\lambda_J)$ defined in \eqref{buildingblock}. Therefore, the leading term of the large $N$  partition function  finally reads

\begin{equation}
Z_{A_{n-1}}(\{ \lambda_I\})=\Big(\prod_{I=1}^n Z_{\mathcal{N}=4}(\lambda_I)\Big) \frac{\mathfrak{F}(\lambda_1,\lambda_2)\cdots \mathfrak{F}(\lambda_n,\lambda_1)}{\mathfrak{F}(\lambda_1,\lambda_1)\cdots \mathfrak{F}(\lambda_n,\lambda_n)}\, .
\label{Zfactorizes}
\end{equation}
This makes the  ``modular" structure of the large $N$ partition function explicit for arbitrary couplings. 

In summary, drawing the quiver in $\mathcal{N}=2$ notation, we can construct the large $N$ partition function using the rules that each node contributes to the partition function a factor  

\begin{equation}
\label{node}
\mathscr{Z}_{\rm node}(\lambda_I)=\frac{Z_{\mathcal{N}=4}(\lambda_I)} 
{\mathfrak{F}(\lambda_I,\lambda_I)}\, ;
\end{equation}
while each link between nodes $I$ and $J$ contributes

\begin{equation}
\label{hyper}
\mathscr{Z}_{\rm link}(\lambda_I,\lambda_{J})=\mathfrak{F}(\lambda_I,\lambda_J)\, .
\end{equation}
Then, the leading contribution to the large $N$ partition function for the $A_{n-1}$ quiver is

\begin{equation}
\label{ZAn-1}
Z_{A_{n-1}}(\{\lambda_I\})=\prod_{I=1}^n \mathscr{Z}_{\rm node}(\lambda_I)
\mathscr{Z}_{\rm link}(\lambda_I,\lambda_{I+1})\, ,
\end{equation}
where the node $n+1$ is identified with the node 1. 

Note that setting all but one $\lambda_I$ to zero we recover the $\mathcal{N}=2$  SCQCD  case, so that

\begin{equation}
\label{Z2-Z4}
Z_{\rm \small SCQCD}(\lambda_I)=\frac{Z_{\mathcal{N}=4}(\lambda_I)}{\mathfrak{F}(\lambda_I,\lambda_I)}\, .
\end{equation}
%

\subsection{Convergence properties of planar perturbation series}

\paragraph{${\cal N}=2$ SCQCD.}

Consider the free energy $F=-\ln Z$. In the large $N$ limit, we have found the exact formula
\begin{equation}
F= F_{{\cal N}=4} +  \ln {\mathfrak{F}(\lambda,\lambda )}\, ,
\end{equation}
where $F_{{\cal N}=4}=- N^2\ln\sqrt{\lambda}$ and 
the explicit expression of $\mathfrak{F}(\lambda,\lambda )$ can be read from (\ref{buildingblock}). We get
\begin{equation}
 \ln {\mathfrak{F}(\lambda,\lambda )} =N^2 \sum_{n=2}^\infty  \frac{ (-1)^n (2n)!}{n \ 4^n} \left( \frac{\lambda} {4\pi^2}\right)^n \zeta(2n-1) \sum_{k=1}^{n-1} \frac{1}{k!(k+1)!(n-k)!(n-k+1)!}\, .
\end{equation}
Using the formula
\begin{equation}
    \sum_{k=1}^{n-1} \frac{1}{k!(k+1)!(n-k)!(n-k+1)!} =\frac{4^{n+1} \Gamma \left(n+\frac{3}{2}\right)}{\sqrt{\pi } n! (n+1)! (n+2)!}- \frac{2}{n! (n+1)!}\ ,
\end{equation}
we obtain
$$
\ln \mathfrak{F}(\lambda,\lambda ) = N^2(S_1-S_2)\ ,
$$
where
\begin{eqnarray}
S_1 &=& 4\sum_{n=2}^\infty  \frac{ (-1)^n }{n } \frac{(2n)! \Gamma \left(n+\frac{3}{2}\right)}{\sqrt{\pi } n! (n+1)! (n+2)!}
\left( \frac{\lambda} {4\pi^2}\right)^n \zeta(2n-1) \ ,
\nonumber\\
S_2 &=&
2\sum_{n=2}^\infty  \frac{ (-1)^n }{n} \frac{(2n)!}{n! (n+1)!}
\frac{1}{4^n}\left( \frac{\lambda} {4\pi^2}\right)^n \zeta(2n-1) \ .
\end{eqnarray}
Let us consider the convergence properties of these series.
For $n\gg 1$, $ \zeta(2n-1)\to 1$.
Using the de Moivre-Stirling formula for the $\Gamma $ function and factorials, we then easily find that the radius of convergence of
$S_1 $ is $|\lambda|<\pi^2$, while the radius of convergence of $S_2$ is  $|\lambda|<4\pi^2$.
More precisely, the series converges for $|\lambda|\leq \pi^2$, $\lambda\neq -\pi^2$. 
At $\lambda =-\pi^2$, it has a logarithmic branch-point, with behavior
$\sim (\lambda +\pi^2)^3\ln (\lambda+\pi^2)$.

In conclusion, the free energy in ${\cal N}=2$ SCQCD  has radius of convergence $|\lambda| =\pi^2$. A finite radius of convergence is expected in the perturbation
theory of planar Feynman diagrams.
In quantum field theory, the convergence properties are governed by the  combinatorics of Feynmann graphs and typically do not depend on the specific observable.
Thus it is natural to expect that other observables will be given by a planar perturbation series that converges  in the
region $|\lambda|<\pi^2$ (though there may be special observables, {\it e.g.} with high supersymmetry, for which the corresponding planar perturbation series
truncates at some order or has infinite radius of convergence).

\paragraph{${\cal N}=2$ superconformal quivers.}

In this case we need to study the convergence properties of the building block (\ref{buildingblock}).
In what follows we will show that the series converges uniformly for
all $\lambda_1, \ \lambda_2$ with
$|\lambda_1|<\pi^2$, $|\lambda_2|<\pi^2$.
As the expression (\ref{buildingblock}) is symmetric under $\lambda_1\leftrightarrow \lambda_2$, we only need to look at the interval  $0<|\lambda_2|/|\lambda_1|<1$.
In this case, it is convenient to write it in the form:
\begin{equation}
\label{sumaq} \ln {\mathfrak{F}}(\lambda_1,\lambda_2)  =N^2 \sum_{n=2}^\infty  \frac{ (-1)^n (2n)!}{n \ 4^n} \left( \frac{\lambda_1} {4\pi^2}\right)^n \zeta(2n-1) c_n(\lambda_2/\lambda_1)
\end{equation}
 with
\begin{equation}
 c_n(\lambda_2/\lambda_1)\equiv \sum_{k=1}^{n-1} \frac{1}{k!(k+1)!(n-k)!(n-k+1)}\left(\frac{\lambda_2}{\lambda_1}\right)^k\, .
\end{equation}
 The coefficient $c_n(x)$ satisfies the  bound
 $c_n(x)< c_n(1)$ in the interval $x\in [0,1)$.
 Indeed, the case with least radius of convergence occurs when $\lambda_2=\lambda_1$, for which the previous results apply and we find the condition $|\lambda_1|<\pi^2$, $|\lambda_2|<\pi^2$. 
One can check that for $|\lambda_2|\ll |\lambda_1|$ the series converges more rapidly. In this case  one can look at the $k=1$ term, which  goes as $1/(n! (n-1)!)$. The resulting series converges for  $|\lambda_1|<4\pi^2$.

In conclusion, the free energy for all $A_{n-1}$ quiver models uniformly converges under the (necessary and sufficient) condition,  $|\lambda_I|\leq \pi^2$,
$\lambda_I\neq -\pi^2 $, $I=1,...,n$.

It is interesting to note that the same radius of convergence arises   in  the calculation of anomalous dimensions of local operators in ${\cal N}=4$ SYM.
This is seen from the dispersion relation of the magnon excitations of the spin chain \cite{Beisert:2004hm}, which has a square-root branch point at
$\lambda =-\pi^2$. Interestingly, the same radius of convergence also appears in the calculation of the free energy and circular Wilson loop in  ${\cal N}=2^*$ theory obtained by perturbing  ${\cal N}=4$ SYM by a small mass term for the hypermultiplet \cite{Russo:2013kea}. In this case the free energy and the Wilson loop have both a logarithmic branch-point at $\lambda =-\pi^2$.

\subsection{Holography}

It is clear that the same factorization (\ref{Zfactor2}), (\ref{Zfactorizes})
applies at strong coupling ($\lambda_I\gg 1$), since the factor $\mathfrak{F}$ can be computed by resumming the perturbation series in the regime
$|\lambda_I |< \pi^2 $ and then analytically continuing to $\lambda_I> \pi^2$.

Let us consider the particular case when $\lambda_I=\lambda$ for all $I$. Then

\begin{equation}
Z_{A_{n-1}}(\lambda )=\Big(Z_{\mathcal{N}=4}(\lambda)\Big)^n\, ;
\end{equation}
thus recovering the result in \cite{Azeyanagi:2013fla}. 

This case, being insensitive on the resummation in $\mathfrak{F}$, can be extrapolated to strong coupling in a straightforward way. 
It follows that
\begin{equation}
\label{localFQ}
F_{A_{n-1}}=n F_{\mathcal{N}=4}= - n N^2 \ln \sqrt{\lambda}\, ,
\end{equation}
where we have used ({\it c.f.} \eqref{ZN4})
\be
F_{\mathcal{N}=4}=-\ln Z_{\mathcal{N}=4}(\lambda)= - N^2 \ln \sqrt{\lambda}\ .
\label{free4}
\ee
The holographic interpretation of this formula in the $\mathcal{N}=4$ case was discussed in
section 3 of  \cite{Russo:2012ay}. The holographic free energy is given by the on-shell supergravity action on
$AdS_5\times S^5$, supplied by boundary counterterms to cancel quartic, quadratic and a logarithmic divergence, 
$$
F_{\rm sugra}(AdS_5\times S^5) =-\frac{\pi L^3}{2G_N} \ln \frac{L}{r_0}\ ,
$$
where $L$ is the radius of $AdS_5$ and of $S^5$, $r_0$ is a UV short-distance cutoff and $G_N$ is the five-dimensional Newton constant. The logarithmic divergence leads to the trace anomaly
in the energy-momentum tensor.
The coefficient $\frac{\pi L^3}{2G_N}$ is identified with $N^2$ by the AdS/CFT dictionary.
In the gauge theory, the logarithmic divergence arises with the same coefficient proportional to $N^2$ by explicitly computing the one-loop partition
function on $\mathbb{S}^4$ \cite{burgess}.
 A complete comparison of the formulas requires relating the gauge-theory momentum cutoff $\Lambda_0$ and supergravity UV cutoff $r_0$.
This has been done in  \cite{Russo:2012ay} by using a physical argument that identifies $\Lambda_0$ to the maximum possible mass in $AdS_5$ corresponding to a string stretched from the horizon to the cutoff surface. 
One finds $2\pi R\Lambda_0 = \sqrt{\lambda} r_0/L$, where we have re-introduced the $\mathbb{S}^4$ radius $R$. 
In this way one gets a perfect match between (\ref{free4}) and the supergravity formula. 

The identification $2\pi R\Lambda_0 = \sqrt{\lambda} r_0/L$  holds true in the
quiver case, since the argument of \cite{Russo:2012ay} only involves the $AdS_5$ part.
In the quiver case, the calculation on the supergravity side is similar, with the only difference that ${\rm Vol}(S^5/\mathbb{Z}_n)/{\rm Vol}(S^5)=1/n$.
This implies an additional factor $n$ in the five-dimensional Newton constant, leading to 
$$
F_{\rm sugra}( AdS_5\times S^5/\mathbb{Z}_n)=- nN^2 \ln \frac{L}{r_0}=-nN^2 \ln  \sqrt{\lambda} \ .
$$
in agreement with the localization formula (\ref{localFQ}).

\section{Two-point correlation functions for chiral primary operators}\label{sec2}

Let us now turn to the computation of extremal correlation functions in $\mathcal{N}=2$ necklace quiver theories of the form $\langle O(x)\overline{O}'(y)\rangle$, where both $O(x)$ and $O'(y)$ are CPO's. Note that conformal invariance determines that the spacetime dependence of the correlator to be $|x-y|^{-2\Delta_O}\delta_{\Delta_O,\Delta_{O'}}$. Thus, in the following, we will omit such dependence.   

In particular, we will be interested on extremal correlation functions of CPO's in $\mathcal{N}=2$ theories at large $N$. Because of this, the basis of CPO's dramatically simplifies and we just need to consider single-trace operators (see \cite{Rodriguez-Gomez:2016ijh,Baggio:2016skg} for recent discussions in this context). For the present quiver theories, the set of CPO's will consist of trace of powers of the scalars in the vector multiplets. Explicitly, the operators of interest are $X_{n}^I={\rm Tr}X_{I,I}^n$. 
The crucial subtlety is that in the $\mathbb{S}^4$, due to the conformal anomaly, CPO's of different dimensions can mix. Such mixture must be disentangled through a Gram-Schmidt procedure as described in \cite{Gerchkovitz:2016gxx}. 
Then, in general, the $\mathbb{R}^4$ operator $X_{n}$ corresponds to a linear combination of the $\mathbb{S}^4$ operators as

\begin{equation}
\label{mixing}
X_{n}^I=\sum_{m\leq n} c_{J,n}^{I,m} X_{m}^J\Big|_{\mathbb{S}^4}\, ,
\end{equation}
where the operators $X_{n}^I\Big|_{\mathbb{S}^4}$ are given by $X_{n}^I\Big|_{\mathbb{S}^4}=\sum_i (a_i^I)^n$, with $a_i^I$ being the real part of the eigenvalues of the scalar in the $I$'th vector multiplet. These are nothing but the coordinates of the Cartan torus of the gauge group over which the partition function integrates. As described, the mixing coefficients $c_{J,n}^{I,m}$ are to be determined through the Gram-Schmidt procedure. Note that $c_{I,n}^{J,n}=\delta_I^J$. Then, the correlators on $\mathbb{R}^4$ are

\begin{equation}
\label{correlatorformal}
\langle X_n^I  \overline{X}_{n'}^{J}\rangle = \sum_{m,m'} c_{L,n}^{I,m}c_{L',n'}^{J,m'} \langle\langle X_m^L\Big|_{\mathbb{S}^4}\overline{X}_{m'}^{L'}\Big|_{\mathbb{S}^4}\rangle\rangle\, ;
\end{equation}
where $\langle\langle A B \rangle\rangle$ stands for the correlator in the  matrix model for the theory on $\mathbb{S}^4$. Note that inside $\langle\langle\cdot\rangle\rangle$ only operators in the $\mathbb{S}^4$ matrix model can appear. Thus, we can omit the $\mathbb{S}^4$ label at no risk of confusion. Moreover, in the $\mathbb{S}^4$ matrix model $X^I$ becomes purely real, and thus the hermitean conjugation can be dropped as well.

From \eqref{correlatorformal} we see that the quantities of primary interest are the two-point functions in the sphere matrix model. These are given by

\begin{equation}
\label{S4correlatorGENERIC}
\langle \langle X_n^I X_m^J\rangle\rangle=Z_{A_{n-1}}^{-1}  \int \prod_{I=1}^n\left( d^{N-1} a^I_i \,\Delta_I\ e^{- \frac{8\pi^2}{g_I^2}\sum_{i=1}^N (a^I_i)^2}\right) \, Z_{\rm 1-loop}\,Z_{\rm inst} \, \Big(\sum (a_i^I)^n\Big)\,\Big(\sum (a_i^J)^m\Big)\,\, .
\end{equation}

Note that in the mapping between $\mathbb{R}^4$ and $\mathbb{S}^4$ operators defined in \eqref{mixing} there is a contribution --only non-vanishing for even $n$-- of the identity operator. This particular mixing is easily taken into account by slightly modifying \eqref{mixing} as

\begin{equation}
\label{mixingVEVless}
X_n^I=\sum_{m\leq n} \hat{c}_{J,n}^{I,m} \mathscr{X}_m^J\, ,
\end{equation}
where $\mathscr{X}_m^J=X_m^J-\langle\langle X_m^J\rangle\rangle$. For these we have

\begin{equation}
\label{Obonitacorrelators}
\langle\langle \mathscr{X}_{n}^I \mathscr{X}_{m}^{J}\rangle\rangle=\langle \langle X_n^I X_m^J\rangle\rangle-\langle \langle X_n^I\rangle\rangle\langle\langle X_m^J\rangle\rangle\, ;
\end{equation}
so that the correlators of interest are given by

\begin{equation}
\label{correlatorformalmathscrcX}
\langle X_n^I  \overline{X}_{n'}^{J}\rangle = \sum_{m,m'} \hat{c}_{L,n}^{I,m}\hat{c}_{L',n'}^{J,m'} \langle\langle \mathscr{X}_m^L\mathscr{X}_{m'}^{L'}\rangle\rangle\, ,
\end{equation}
where the $\hat{c}_{L,n}^{I,m}$ are to be fixed through the Gram-Schmidt procedure as in \cite{Gerchkovitz:2016gxx}. For clarity, we summarize the notation used throughout the paper in appendix \ref{Notation}. Note that \eqref{Obonitacorrelators} is nothing but the corresponding connected correlator in the quiver theory.

In the following, using \eqref{S4correlatorGENERIC}, we will compute  \eqref{correlatorformalmathscrcX} at large $N$.
Just as in the earlier discussion of the partition function, in the large $N$ limit the instanton factor $Z_{\rm inst}$ in \eqref{S4correlatorGENERIC}
will be set to 1. 
Moreover, we will consider the weak coupling regime where we can use the expansion of  \eqref{expansionH} to compute, perturbatively, the  correlators
on $\mathbb{S}^4$.

\subsection{Correlators in $\mathcal{N}=2$ superconformal QCD}\label{secQCD}

Two-point functions in $\mathcal{N}=2$ superconformal QCD at large $N$ have been computed in \cite{Rodriguez-Gomez:2016ijh} including the terms with coefficients proportional to $\zeta(3)$ and $\zeta(3)^2$. Our first task will be to improve these results by computing all the terms in the planar expansion that have coefficients with linear dependence on the $\zeta$'s (as well as the $\zeta(3)^2$ term).

In the case of $\mathcal{N}=2$ SCQCD we just have a single gauge group. Denoting the scalar in the vector multiplet by $X$, the CPO's of interest are $X_n\equiv {\rm Tr}X^n$. These CPO's  are represented in $\mathbb{S}^4$ by $X_n=\sum_i a_i^n$, where $a_i$ are the variables of integration in the $\mathbb{S}^4$ partition function. As explained above (see appendix \ref{Notation} for a summary of the notation), $\langle X_n \bar X_m\rangle$ will denote the correlators for the theory on $\mathbb{R}^4$, while $\langle\langle X_n\overline{X}_m\rangle\rangle$ will denote correlators for the theory on $\mathbb{S}^4$. 
Moreover, $\mathscr{X}_n$ will refer to VEV-less operators in $\mathbb{S}^4$, whose two-point correlators are nothing but the connected two-point correlators in the $\mathbb{S}^4$ matrix model.

By expanding the one-loop determinant we can compute the connected correlator in $\mathcal{N}=2$ SCQCD in the planar perturbative expansion in powers of $\lambda$. In general, this will include an infinite series of corrections to the $\mathcal{N}=4$ correlator with coefficients proportional to products of $\zeta$ functions. 
We can unambiguously isolate and compute the contribution of the terms with linear
dependence on $\zeta's$.
In particular, they could be compared with similar terms that should appear in
a direct Feynman diagram calculation.
Thus, including all terms linear in $\zeta$ as well as the first $\zeta^2$ term proportional to $\zeta(3)^2$, the result is (we compile the details in appendix \ref{correlatorsN=2})

\begin{equation}
\label{N=2}
\langle\langle \mathscr{X}_n\mathscr{X}_m\rangle\rangle=\mathscr{C}_{n,m} -\sum_{q=2}^\infty\sum_{k=2}^{2q-2} (-1)^{q+k}\frac{\zeta(2q-1)}{q} \binom{2q}{k} M^{(1)}_{k,q,n,m}+\frac{9}{2}\zeta(3)^2\,M^{(2)}\, ,
\end{equation}
where, to leading order in $N$, we have defined

\begin{equation}
M^{(1)}_{k,q,n,m}= \mathscr{C}_{k,n,m}\mathscr{C}_{2q-k} + \mathscr{C}_{2q-k,n,m}\mathscr{C}_{k}  +\mathscr{C}_{k,n} \mathscr{C}_{2q-k,m} +\mathscr{C}_{2q-k,n} \mathscr{C}_{k,m}\, ;
\end{equation}
and

\begin{equation}
 M^{(2)}=4\mathscr{C}_2\mathscr{C}_2 \mathscr{C}_{2,2,n,m}+ 8 \mathscr{C}_2  \mathscr{C}_{2,2,n} \mathscr{C}_{2,m} + 8 \mathscr{C}_2 \mathscr{C}_{2,2,m} \mathscr{C}_{2,n}+ 8 \mathscr{C}_{2,2} \mathscr{C}_{2,m} \mathscr{C}_{2,n} +8\mathscr{C}_2 \mathscr{C}_{2,2} \mathscr{C}_{2,n,m}  \, .
\end{equation}

For concreteness, let us concentrate on correlators for even operators. For them, in order to have a non-vanishing result, $k$ in $M^{(1)}$ must be even. 
Then, using the results for $\mathscr{C}_{2m,2n,2r}$ and $\mathscr{C}_{2m,2n,2r,2s}$ 
given in appendix \ref{correlatorsN=4},
we can easily compute the connected correlators $\langle\langle \mathscr{X}_n\mathscr{X}_m\rangle\rangle$ and perform the Gram-Schmidt orthogonalization to find the correlators for the theory on $\mathbb{R}^4$. Explicitly, up to $\zeta(13)$, for the first few dimensions we find

\begin{eqnarray}
\label{SCQCD :22}
\langle X_2\overline{X}_2\rangle&=&\frac{2\lambda^2}{(2\pi)^4}\Big[1-\frac{9}{4}\zeta(3)\Big(\frac{\lambda}{4\pi^2}\Big)^2+\frac{15}{2}\zeta(5)\Big(\frac{\lambda}{4\pi^2}\Big)^3 +\Big(\frac{45}{8}\zeta(3)^2 -\frac{175}{8}\zeta(7)\Big)\Big(\frac{\lambda}{4\pi^2}\Big)^4\nonumber \\  && \hspace{1cm} +\frac{4095}{64}\zeta(9)\Big(\frac{\lambda}{4\pi^2}\Big)^5  -\frac{98637}{512}\zeta(11)\Big(\frac{\lambda}{4\pi^2}\Big)^6+\frac{153153}{256}\zeta(13)\Big(\frac{\lambda}{4\pi^2}\Big)^7+\cdots\Big] \nonumber \\
\end{eqnarray}
\begin{eqnarray}
\label{SCQCD :44}
\langle X_4\overline{X}_4\rangle &=& \frac{4\lambda^4}{(2\pi)^8}\Big[1-3\zeta(3)\Big(\frac{\lambda}{4\pi^2}\Big)^2+10\zeta(5)\Big(\frac{\lambda}{4\pi^2}\Big)^3 +\Big( \frac{63}{8}\zeta(3)^2-\frac{1855}{64}\zeta(7)\Big)\Big(\frac{\lambda}{4\pi^2}\Big)^4\nonumber \\  && \hspace{1cm} +\frac{2709}{32}\zeta(9)\Big(\frac{\lambda}{4\pi^2}\Big)^5  -\frac{16401}{64}\zeta(11)\Big(\frac{\lambda}{4\pi^2}\Big)^6+\frac{6435}{8}\zeta(13)\Big(\frac{\lambda}{4\pi^2}\Big)^7+\cdots\Big] \nonumber \\
\end{eqnarray}
\begin{eqnarray}
\label{SCQCD :66}
\langle X_6\overline{X}_6\rangle&=&\frac{6\lambda^6}{(2\pi)^{12}}\Big[1-\frac{9}{2}\zeta(3)\Big(\frac{\lambda}{4\pi^2}\Big)^2+15\zeta(5)\Big(\frac{\lambda}{4\pi^2}\Big)^3+\Big(\frac{243}{16}\zeta(3)^2 -\frac{1365}{32}\zeta(7)\Big) \Big(\frac{\lambda}{4\pi^2}\Big)^4\nonumber \\  && \hspace{1cm} +\frac{7749}{64}\zeta(9)\Big(\frac{\lambda}{4\pi^2}\Big)^5  -\frac{181797}{512}\zeta(11)\Big(\frac{\lambda}{4\pi^2}\Big)^6+\frac{276705}{256}\zeta(13)\Big(\frac{\lambda}{4\pi^2}\Big)^7+\cdots\Big] \nonumber \\
\end{eqnarray}
\begin{eqnarray}
\label{SCQCD :88}
\langle X_8\overline{X}_8\rangle&=&\frac{8\lambda^8}{(2\pi)^{16}}\Big[1-6\zeta(3)\Big(\frac{\lambda}{4\pi^2}\Big)^2+20\zeta(5)\Big(\frac{\lambda}{4\pi^2}\Big)^3+\Big(\frac{99}{4}\zeta(3)^2 -\frac{455}{8}\zeta(7)\Big) \Big(\frac{\lambda}{4\pi^2}\Big)^4\nonumber\\  && \hspace{1cm} +\frac{2583}{16}\zeta(9)\Big(\frac{\lambda}{4\pi^2}\Big)^5  -\frac{30261}{64}\zeta(11)\Big(\frac{\lambda}{4\pi^2}\Big)^6+\frac{45903}{32}\zeta(13)\Big(\frac{\lambda}{4\pi^2}\Big)^7+\cdots\Big]\nonumber \\ 
\end{eqnarray}
These expressions contain and extend the ones found in \cite{Rodriguez-Gomez:2016ijh} to higher orders in the planar perturbative expansion.

\subsubsection{Effective couplings for $\mathcal{N}=2$ superconformal QCD}\label{SCQDrep}

It was suggested in \cite{Pomoni:2013poa} that observables in the purely gluonic $SU(2,1|2)$ subsector of any planar $\mathcal{N}=2$ theory can be computed upon performing a coupling replacement in the $\mathcal{N}=4$ result. Elaborating on this, in \cite{Mitev:2014yba} it was shown that the VEV of the supersymmetric circular Wilson loop
in SCQCD and in the $A_{n-1}$ quiver gauge theories 
has exactly the same structure as in the $\mathcal{N}=4$ theory, indeed with the replacement
of the coupling $\lambda $ by a new effective coupling.
A natural question is whether this observation also applies to two-point correlation functions. In other words, whether the two-point correlation functions of all CPO's have the same form as in 
 the $\mathcal{N}=4$ theory, in terms of a new universal, effective coupling, for all correlators. 
  Below we will show that this works for  the leading 
  $\zeta(3) $ and $\zeta(3)^2$ terms, but not in general for the other contributions.

Let us first recall how the results 
of \cite{Mitev:2014yba} arise using the formulas given in the appendix.
For the VEV, we have (for simplicity, here we show only linear terms in the $\zeta$'s)

\begin{equation}
\langle\langle X_n\rangle\rangle = \langle\langle X_n \rangle\rangle_0 -\sum_{q=2}^\infty\sum_{k=1}^{q-1} (-1)^{q}\frac{\zeta(2q-1)}{q} \binom{2q}{2k}M_{k,q,n}\, ,
\end{equation}
with

\begin{equation}
M_{k,q,n}= \mathscr{C}_{2k} \mathscr{C}_{2q-2k,n}+\mathscr{C}_{2q-2k} \mathscr{C}_{2k,n}\, .
\end{equation}
With this, it is easy to compute the VEV of the circular Wilson loop

\begin{equation}
\langle W\rangle =\frac{1}{N}\langle {\rm Tr}e^{-2\pi X}\rangle\, ,
\end{equation}
to all orders for the terms with linear $\zeta $ coefficient, by expanding in powers of $X$.
This agrees with the results in \cite{Mitev:2014yba},  showing that the VEV of the circular Wilson loop is equal to that of $\mathcal{N}=4$ SYM \cite{Drukker:2000rr} upon performing the same coupling replacement given in \cite{Mitev:2014yba,Fraser:2015xha,Mitev:2015oty}.

 Now let us consider two-point correlation functions. 
 Let us denote the required effective coupling in the correlator $\langle X_n\overline{X}_n\rangle$ by $\lambda_{\rm eff}^{(n)}$.
 For the terms with coefficients linear in $\zeta(2n-1)$, if one tries to re-write the correlators for $\langle X_{2n}\overline{X}_{2n}\rangle$ in terms of an effective coupling  on the $\mathcal{N}=4$ SYM result, one finds

\begin{eqnarray}
\lambda_{\rm eff}^{(2)} & = & \lambda\,\Big[ 1-\frac{9}{8}\zeta(3) \Big(\frac{\lambda}{4\pi^2}\Big)^2+\frac{15}{4} \zeta(5) \Big(\frac{\lambda}{4\pi^2}\Big)^3 \\ \nonumber &&-\frac{175}{16} \zeta(7)\Big(\frac{\lambda}{4\pi^2}\Big)^4+\frac{4095}{128}\zeta(9)\Big(\frac{\lambda}{4\pi^2}\Big)^5-\frac{98637}{1024} \zeta(11) \Big(\frac{\lambda}{4\pi^2}\Big)^6
+\cdots \Big]\\
\lambda_{\rm eff}^{(4)} & =& \lambda\,\Big[ 1-\frac{3}{4}\zeta(3) \Big(\frac{\lambda}{4\pi^2}\Big)^2+\frac{5}{2} \zeta(5) \Big(\frac{\lambda}{4\pi^2}\Big)^3 \\ \nonumber &&-\frac{1855}{256} \zeta(7)\Big(\frac{\lambda}{4\pi^2}\Big)^4+\frac{2709}{128}\zeta(9)\Big(\frac{\lambda}{4\pi^2}\Big)^5-\frac{16401}{256} \zeta(11) \Big(\frac{\lambda}{4\pi^2}\Big)^6
+\cdots \Big]\\
\lambda_{\rm eff}^{(6)} & =& \lambda\,\Big[ 1-\frac{3}{4}\zeta(3) \Big(\frac{\lambda}{4\pi^2}\Big)^2+\frac{5}{2} \zeta(5) \Big(\frac{\lambda}{4\pi^2}\Big)^3 \\ \nonumber &&-\frac{455}{64} \zeta(7)\Big(\frac{\lambda}{4\pi^2}\Big)^4+\frac{2583}{128}\zeta(9)\Big(\frac{\lambda}{4\pi^2}\Big)^5-\frac{60599}{1024} \zeta(11) \Big(\frac{\lambda}{4\pi^2}\Big)^6
+\cdots \Big]\\
\lambda_{\rm eff}^{(8)} & =& \lambda\,\Big[ 1-\frac{3}{4}\zeta(3) \Big(\frac{\lambda}{4\pi^2}\Big)^2+\frac{5}{2} \zeta(5) \Big(\frac{\lambda}{4\pi^2}\Big)^3 \\ \nonumber &&-\frac{455}{64} \zeta(7)\Big(\frac{\lambda}{4\pi^2}\Big)^4+\frac{2583}{128}\zeta(9)\Big(\frac{\lambda}{4\pi^2}\Big)^5-\frac{30261}{512} \zeta(11) \Big(\frac{\lambda}{4\pi^2}\Big)^6
+\cdots \Big]
\end{eqnarray}
As it is clear from these expressions, there is no universal effective coupling. It is however interesting to note that $\lambda_{\rm eff}^{(4)}$ and $\lambda_{\rm eff}^{(2)}$ differ at $\mathcal{O}(\zeta(3))$; 
$\lambda_{\rm eff}^{(6)}$ and $\lambda_{\rm eff}^{(4)}$ differ at $\mathcal{O}(\zeta(7))$; $\lambda_{\rm eff}^{(8)}$ and $\lambda_{\rm eff}^{(6)}$ differ at $\mathcal{O}(\zeta(11))$; thus suggesting a pattern such that 

\begin{equation}
\label{replacementdifference}
\lambda_{\rm eff}^{(n+2)}-\lambda_{\rm eff}^{(n)} \sim \mathcal{O}\big(\zeta(2n-1)\lambda^n\big)\, .
\end{equation}
Although we have only shown this property for the terms linear in $\zeta$'s, this is sufficient to conclude that there is no universal effective coupling. 
Of course, to a given order in $\lambda$ there will be contributions not only linear in $\zeta$ but also non-linear, \textit{i.e.} products of $\zeta$'s
that we have not computed. It is nevertheless tempting to conjecture that \eqref{replacementdifference} holds as a full-fledged 't Hooft coupling expansion.

An equivalent way to phrase this result is noticing that the correlator $\langle X_n\overline{X}_n\rangle$ in $\mathcal{N}=2$ SCQCD can be written as that of $\mathcal{N}=4$ theory upon a coupling replacement plus corrections of the form

\begin{eqnarray}
\langle X_n\overline{X}_n\rangle_{\rm \small SCQCD}
&=& \langle X_n\overline{X}_n\rangle_{\mathcal{N}=4}(\lambda_{\rm eff})\ 
\left[ 1 +\mathcal{O}\big(\zeta(2n-1)\lambda^n\big)\right]
\nonumber\\
&=& \frac{n\lambda^n_{\rm eff}}{(2\pi)^{2n}}
\left[ 1 +\mathcal{O}\big(\zeta(2n-1)\lambda^n \big)\right]\, ;
\label{gyb}
\end{eqnarray}
being the effective coupling

\begin{eqnarray}
\lambda_{\rm eff}&=&\lambda\,\Big[ 1-\frac{3}{4}\zeta(3) \Big(\frac{\lambda}{4\pi^2}\Big)^2+\frac{5}{2} \zeta(5) \Big(\frac{\lambda}{4\pi^2}\Big)^3-\frac{455}{64} \zeta(7)\Big(\frac{\lambda}{4\pi^2}\Big)^4  \nonumber \\ && +\frac{2583}{128}\zeta(9)\Big(\frac{\lambda}{4\pi^2}\Big)^5-\frac{30261}{512} \zeta(11) \Big(\frac{\lambda}{4\pi^2}\Big)^6+\cdots]\, .
\end{eqnarray}
In terms of $g^2=\frac{\lambda}{(4\pi)^2}$ this is

\begin{equation}
g^2\rightarrow g^2\Big(1-12 g^4\zeta(3)+160g^6\zeta(5)-1820 g^8 \zeta(7) +20664 g^{10}\zeta(9)-242088g^{12}\zeta(11)+\cdots\Big)\, .
\end{equation}
This replacement differs from that in \cite{Mitev:2014yba,Fraser:2015xha,Mitev:2015oty} already at order $\zeta(5)$.

From the above results, it is clear, for example, that if we
restrict the discussion to correlators $\langle X_n\overline{X}_n\rangle_{\rm \small SCQCD}$ with $n>2$, then up to (and including) the order $\zeta(3)\lambda^2 $ we can describe correlators in terms of an effective coupling. We can test this by also including the term with coefficient $\zeta(3)^2$.
Thus let us compute correlators with $n>2$ by  truncating the series and keeping only  the terms with coefficients proportional to $\zeta(3)$ and $\zeta(3)^2$. Starting with the case of odd $n$, $m$, we find

\begin{equation}
\langle\langle \mathscr{X}_n\mathscr{X}_m\rangle\rangle=\Big[1-\frac{3}{8}\zeta(3) (n+m) \Big( \frac{\lambda}{4\pi^2}\Big)^2 +\frac{9}{128} \zeta(3)^2 (m+n)(6+m+n)\Big( \frac{\lambda}{4\pi^2}\Big)^4\Big]\mathscr{C}_{n,m}
\end{equation}
Here we recognize the first terms in the expansion of

\begin{equation}
\langle\langle \mathscr{X}_n\mathscr{X}_m\rangle\rangle=\Big[1-\frac{3}{4}\zeta(3) \Big( \frac{\lambda}{4\pi^2}\Big)^2 +\frac{9}{8} \zeta(3)^2\Big( \frac{\lambda}{4\pi^2}\Big)^4\Big]^{\frac{n+m}{2}}\mathscr{C}_{n,m}
\end{equation}
In turn, since the $\mathscr{C}_{n,m}$ are nothing but the correlators in $\mathcal{N}=4$ theory, proportional to $\lambda^{\frac{n+m}{2}}$, we see that the result in $\mathcal{N}=2$ superconformal QCD up to the order we are discussing is akin to taking the $\mathcal{N}=4$  correlators on $\mathbb{S}^4$ and perform the substitution 

\begin{equation}
\label{subs}
\lambda\rightarrow \lambda_{\rm eff}= \lambda \, \Big(1-\frac{3}{4}\zeta(3) \Big( \frac{\lambda}{4\pi^2}\Big)^2 +\frac{9}{8} \zeta(3)^2\Big( \frac{\lambda}{4\pi^2}\Big)^4\Big)\, .
\end{equation}
Moreover, since the substitution is independent on $n$, $m$, the Gram-Schmidt procedure can be directly imported from the $\mathcal{N}=4$ case. Thus, the result for $\mathbb{R}^4$ correlators is finally identical to that of $\mathcal{N}=4$ theory -- given, for instance, in eq. (3.54) in \cite{Rodriguez-Gomez:2016ijh}-- with the substitution \eqref{subs}.

In terms of $g^2=\frac{\lambda}{(4\pi)^2}$, this substitution corresponds to

\begin{equation}
\label{subsg}
g^2\rightarrow g^2(1-12g^4\zeta(3)+288g^8\zeta(3)^2)\, .
\end{equation}
This is the same coupling replacement found in \cite{Mitev:2014yba,Fraser:2015xha,Mitev:2015oty}.

Let us now consider two-point correlation functions of even operators.
For $n,m\geq 2$, we now find

\begin{equation}
\langle\langle \mathscr{X}_n\mathscr{X}_m\rangle\rangle=\Big[1-3\zeta(3)\Big( \frac{\lambda}{4\pi^2}\Big)^2 \frac{(m+n)(12+2m+2n+m n)}{8(2+m)(2+n)}\Big]\mathscr{C}_{n,m}\, ,
\end{equation}
which recovers the same formula given in eq. (4.32) in \cite{Rodriguez-Gomez:2016ijh} (to compare, one should replace above  $(n,m)\rightarrow (2n,2m)$). Thus, we can again import the Gram-Schmidt orthogonalization from the $\mathcal{N}=4$ case leading to the final result in eq. (4.40) in \cite{Rodriguez-Gomez:2016ijh}. 
The conclusion is as follows.
 For $n>2$, the correlator $\langle X_n\overline{X}_n\rangle$ in $\mathcal{N}=2$ SCQCD, including the terms with coefficients $\zeta(3)$ and $\zeta(3)^2$, given by eq. (4.40) in \cite{Rodriguez-Gomez:2016ijh}, can also be obtained by taking the $\mathcal{N}=4$ result upon performing the substitution \eqref{subs}. 
 On the other hand, the correlator for $\langle X_2\overline{X}_2\rangle$ given in eq. (4.40) of  \cite{Rodriguez-Gomez:2016ijh} (which, it should be stressed, is in agreement with previous results in the literature \cite{Baggio:2015vxa}, \cite{Baggio:2014ioa},\cite{Baggio:2014sna},\cite{Gerchkovitz:2016gxx}) does not follow this pattern, as expected, given the  formula (\ref{gyb}).


\subsection{The quiver theory case}\label{secquiver}

For general $A_{n-1}$ quivers, the correlators of interest are $\langle\langle \mathscr{X}^L_n \mathscr{X}^L_m\rangle\rangle$, $\langle\langle \mathscr{X}^L_n \mathscr{X}^{L+1}_m\rangle\rangle$ and $\langle\langle \mathscr{X}^L_n \mathscr{X}^M_m\rangle\rangle$ where $|L-M|>1$. 

It is useful to explicitly show the 't Hooft coupling dependence in the correlators. To this purpose, we define

\begin{equation}
\mathscr{C}^L_{m_1,\cdots,m_n}=\lambda_L^{\frac{m_1+\cdots+m_n}{2}}\hat{\mathscr{C}}_{m_1,\cdots,m_n}\, ,
\end{equation}
where $\hat{\mathscr{C}}_{m_1,\cdots,m_n}$ is a copy of $\mathscr{C}^L_{m_1,\cdots,m_n}$ evaluated at $\lambda_L=1$.
Then, following the same strategy as for $\mathcal{N}=2$ SCQCD , we find (see appendix \ref{correlatorsquiver} for details of the computation)

\begin{eqnarray}
&&\nonumber \langle\langle \mathscr{X}^L_{n} \mathscr{X}^L_{m}\rangle\rangle=\langle\langle \mathscr{X}^L_{n} \mathscr{X}^L_{m}\rangle\rangle_{\rm \small SCQCD}(\lambda_L) 
\nonumber\\
&&+\sum_{q=2}^\infty\sum_{k=1}^{q-1}\frac{(-1)^{q}}{q}\zeta(2q-1)\binom{2q}{2k}\Big(
\lambda^k_{L+1}\lambda_L^{\frac{m+n}{2}+q-k}\hat{\mathscr{C}}_{2k}\hat {\mathscr{C}}_{2q-2k,n,m} +\lambda^{q-k}_{L-1}\lambda_L^{\frac{m+n}{2}+k}\hat{\mathscr{C}}_{2q-2k}\hat{\mathscr{C}}_{2k,n,m}\Big)\, ;\nonumber \\ && \\ 
&&\langle\langle \mathscr{X}^L_{n} \mathscr{X}^{L+1}_{m}\rangle\rangle = \sum_{q=2}^\infty\sum_{k=1}^{2q-1}\frac{(-1)^{q+k}}{q}\zeta(2q-1)\binom{2q}{k}
\lambda^{\frac{m+k}{2}}_{L+1}\lambda_L^{\frac{n+2q-k}{2}}\hat{\mathscr{C}}_{m,k}\hat{\mathscr{C}}_{2q-k,n}\, ; \\
&&\langle\langle \mathscr{X}^L_{n} \mathscr{X}^M_{m}\rangle\rangle = 0\,, \qquad |L-M|>1\, .
\end{eqnarray}
The coefficients $\hat{\mathscr{C}}_{m_1,\cdots,m_n}$ are given in terms of factorials (see appendix B).

\subsection{The $A_1$ quiver}\label{sect4}



In the case of the $A_1$ theory, we have two sets of VEV-less operators in the $\mathbb{S}^4$ matrix model $\mathscr{X}^1_n\equiv\mathscr{X}_n$, $\mathscr{X}_n^2\equiv \mathscr{Y}_n$. 
They are constructed in terms of the scalar fields in the $\mathcal{N}=2$ vector multiplets of the two gauge groups
$SU(N)\times SU(N)$. To begin with, note that the $A_1$ case is  special, as, to a given node, both nearest neighbour nodes are the same. Thus, the $\langle\langle \mathscr{X}_n\mathscr{Y}_m\rangle\rangle$ correlator must be computed separately for this case (see appendix \ref{correlatorsquiver}). We  find \footnote{From the formulas of appendix \ref{correlatorsquiver}, see e.g. \eqref{A1XY}, we extract the coupling dependence and re-write the correlator in terms of the hatted functions.}

\begin{eqnarray}
\nonumber \langle\langle \mathscr{X}_n \mathscr{X}_m\rangle\rangle &=&\lambda_1^{\frac{n+m}{2}}\Big[ \lambda_1^{-\frac{n+m}{2}} \langle\langle \mathscr{X}_n \mathscr{X}_m\rangle\rangle_{\rm \small SCQCD} \\ \nonumber &&  +\sum_{q=2}^\infty \sum_{k=0}^{2q}\frac{(-1)^{q+k}}{q}\zeta(2q-1)\binom{2q}{k}\Big(\lambda_2^{\frac{k}{2}}\lambda_1^{\frac{2q-k}{2}} \hat{\mathscr{C}}_{k}\hat{\mathscr{C}}_{2q-k,n,m} +\lambda_1^{\frac{k}{2}}\lambda_2^{\frac{2q-k}{2}} \hat{\mathscr{C}}_{2q-k}\hat{\mathscr{C}}_{k,n,m}\Big)\Big] \, , \\
\nonumber \langle\langle \mathscr{Y}_n \mathscr{Y}_m\rangle\rangle &=&\lambda_2^{\frac{n+m}{2}}\Big[ \lambda_2^{-\frac{n+m}{2}} \langle\langle \mathscr{Y}_n \mathscr{Y}_m\rangle\rangle_{\rm \small SCQCD} \\ \nonumber && +\sum_{q=2}^\infty \sum_{k=0}^{2q}\frac{(-1)^{q+k}}{q}\zeta(2q-1)\binom{2q}{k}\Big(\lambda_1^{\frac{k}{2}}\lambda_2^{\frac{2q-k}{2}} \hat{\mathscr{C}}_{k}\hat{\mathscr{C}}_{2q-k,n,m} +\lambda_2^{\frac{k}{2}}\lambda_1^{\frac{2q-k}{2}} \hat{\mathscr{C}}_{2q-k}\hat{\mathscr{C}}_{k,n,m}\Big)\Big] \, , \\ 
\nonumber \langle\langle \mathscr{X}_n\mathscr{Y}_m\rangle\rangle &=&\lambda_1^{\frac{n}{2}}\lambda_2^{\frac{m}{2}} \sum_{q=2}^\infty \sum_{k=0}^{2q}\frac{(-1)^{q+k}}{q}\zeta(2q-1)\binom{2q}{k} \Big(\lambda_1^{\frac{2q-k}{2}}\lambda_2^{\frac{k}{2}}\hat{\mathscr{C}}_{2q-k,n}\hat{\mathscr{C}}_{k,m} +\lambda_2^{\frac{2q-k}{2}}\lambda_1^{\frac{k}{2}} \hat{\mathscr{C}}_{2q-k,m}\hat{\mathscr{C}}_{k,n} \Big)\, .\\
\end{eqnarray}

We now define the untwisted $\mathscr{U}_n$ and twisted $\mathscr{T}_n$ operators as

\begin{equation}
\mathscr{U}_n=\frac{\lambda_1^{-\frac{n}{2}}}{\sqrt{2}}\mathscr{X}_n+\frac{\lambda_2^{-\frac{n}{2}}}{\sqrt{2}}\mathscr{Y}_n\,,\qquad \mathscr{T}_n=\frac{\lambda_1^{-\frac{n}{2}}}{\sqrt{2}} \mathscr{X}_n-\frac{\lambda_2^{-\frac{n}{2}}}{\sqrt{2}} \mathscr{Y}_n\, .
\end{equation}
These operators have well defined transformation properties under the $\mathbb{Z}_2$ action which exchanges the two gauge groups,
i.e. $\mathscr{U}_n\to \mathscr{U}_n$ and $\mathscr{T}_n\to - \mathscr{T}_n$ (see also \cite{Gukov:1998kk}).
Note that the perturbative orbifold corresponds to setting $\lambda_1=\lambda_2=\lambda$. On the other hand, the limit of one coupling going to zero gives us back  the $\mathcal{N}=2$ SCQCD case. Then

\begin{equation}
\label{UT}
\langle\langle \mathscr{U}_n\mathscr{T}_m\rangle\rangle=-\sum_{q=2}^\infty\sum_{k=0}^{2q}\frac{(-1)^{q+k}}{2q}\zeta(2q-1)\binom{2q}{k}   \Big[\Big(\lambda_1^q -\lambda_2^q\Big) \hat{M}^{(1)}-\Big(\lambda_2^{\frac{k}{2}}\lambda_1^{\frac{2q-k}{2}} -\lambda_1^{\frac{k}{2}}\lambda_2^{\frac{2q-k}{2}}\Big) \hat{M}_{k,q,n,m}'^{(1)}\Big]
\end{equation}
with

\begin{equation}
\hat{M}_{k,q,n,m}'^{(1)}=\hat{\mathscr{C}}_{k}\hat{\mathscr{C}}_{2q-k,n,m} -\hat{\mathscr{C}}_{2q-k,n}\hat{\mathscr{C}}_{k,m} -\hat{\mathscr{C}}_{2q-k}\hat{\mathscr{C}}_{k,n,m} +\hat{\mathscr{C}}_{2q-k,m}\hat{\mathscr{C}}_{k,n}\, .
\end{equation}
In turn, using the explicit form of the $\mathcal{N}=2$ SCQCD correlator in  \eqref{N=2}, we have

\begin{equation}
\label{UU}
\langle\langle \mathscr{U}_n\mathscr{U}_m\rangle\rangle=\hat{\mathscr{C}}_{n,m} - \sum_{q=2}^\infty \sum_{k=0}^{2q} (-1)^{q+k}\frac{\zeta(2q-1)}{2q} \binom{2q}{k} \Big[ \Big(\lambda_1^q+\lambda_2^q\Big) - \Big(\lambda_2^{\frac{k}{2}} \lambda_1^{\frac{2q-k}{2}}+ \lambda_1^{\frac{k}{2}}\lambda_2^{\frac{2q-k}{2}}\Big)\Big]\hat{M}^{(1)}_{k,q,n,m}\, ;
\end{equation}
where $\hat{M}^{(1)}_{k,q,n,m}$ stands for $M^{(1)}_{k,q,n,m}$ evaluated at coupling one and

\begin{eqnarray}
&& \langle\langle \mathscr{T}_n\mathscr{T}_m\rangle\rangle=\hat{\mathscr{C}}_{n,m} 
\nonumber\\
&&- \sum_{q=2}^\infty \sum_{k=0}^{2q} (-1)^{q+k}\frac{\zeta(2q-1)}{2q} \binom{2q}{k} \Big[\Big(\lambda_1^q+\lambda_2^q\Big)\hat{M}^{(1)}_{k,q,n,m} - \Big(\lambda_2^{\frac{k}{2}} \lambda_1^{\frac{2q-k}{2}}+ \lambda_1^{\frac{k}{2}}\lambda_2^{\frac{2q-k}{2}}\Big)\hat{M}''^{(1)}_{k,q,n,m}\Big]\, ;\nonumber \\
\label{TT}
\end{eqnarray}
being

\begin{equation}
\hat{M}''^{(1)}_{k,q,n,m}= \hat{\mathscr{C}}_{k,n,m}\hat{\mathscr{C}}_{2q-k} + \hat{\mathscr{C}}_{2q-k,n,m}\hat{\mathscr{C}}_{k}  -\hat{\mathscr{C}}_{k,n} \hat{\mathscr{C}}_{2q-k,m} -\hat{\mathscr{C}}_{2q-k,n} \hat{\mathscr{C}}_{k,m}\, .
\end{equation}

\subsubsection{The case $\lambda_1=\lambda_2$}

To begin with, since at the perturbative orbifold both couplings are equal, there is no need to introduce the extra coupling factors in the definition of $\mathscr{U}_n$ and $\mathscr{T}_n$ to preserve the $\mathbb{Z}_2$ symmetry. In fact, in this case it will prove more useful to simply define $\mathscr{U}_n=2^{-\frac{1}{2}}(\mathscr{X}_n+\mathscr{Y}_n)$, $\mathscr{T}_n=2^{-\frac{1}{2}}(\mathscr{X}_n-\mathscr{Y}_n)$ (that this is equivalent to multiplication by the appropriate factor of $\lambda=\lambda_1=\lambda_2$ the above formulas). 

In the perturbative orbifold limit the two towers $\mathscr{U}_n$ and $\mathscr{T}_n$ decouple, as they become orthogonal due to \eqref{UT}. Then, notably, at the perturbative orbifold point $\lambda_1=\lambda_2=\lambda$ the correlator for the untwisted fields $\mathscr{U}_n$ becomes identical to that of $\mathcal{N}=4$ SYM, while that for the twisted fields becomes 

\begin{equation}
\label{TTorbifold}
\langle\langle \mathscr{T}_n\mathscr{T}_m\rangle\rangle=\lambda^{\frac{n+m}{2}}\hat{\mathscr{C}}_{n,m} -2\,\lambda^{\frac{n+m}{2}} \sum_{q=2}\sum_{k=0}^{2q} (-1)^{q+k}\frac{\zeta(2q-1)}{q} \binom{2q}{k} \lambda^q\Big( \hat{\mathscr{C}}_{k,n}\hat{\mathscr{C}}_{2q-k,m} + \hat{\mathscr{C}}_{2q-k,n}\hat{\mathscr{C}}_{k,m} \Big)\, .
\end{equation}
One can then perform the Gram-Schmidt orthogonalization for the tower of untwisted fields (this is just identical to the $\mathcal{N}=4$ case in \cite{Rodriguez-Gomez:2016ijh}) finding

\begin{equation}
\label{UUgeneral}
\langle U_n\overline{U}_n\rangle = \frac{n \lambda^n}{(2\pi)^{2n}}\, .
\end{equation}
This result might be compared with the dual holographic computation in \cite{Lee:1998bxa}, where
 2-point functions for $\mathcal{N}=4$ CPO's are computed using holography. In turn, $\mathcal{N}=4$ SYM is the worldvolume theory on D3 branes on $\mathbb{C}^3$ and the three chiral multiplets correspond to each $\mathbb{C}$ plane. In $\mathcal{N}=2$ notation, one of them is part of the vector multiplet, and this is the one whose correlators are computed in \cite{Lee:1998bxa}. In constructing the quiver theory, the orbifold can be taken to act on the transverse $\mathbb{C}^2$. Thus, the computation in \cite{Lee:1998bxa} should go essentially unchanged for the untwisted sector, thus leading to the result \eqref{UUgeneral}.

In turn, running the Gram-Schmidt for the twisted sector operators, we find

\begin{eqnarray}
\label{TT:22}
\langle T_2\overline{T}_2\rangle&=&\frac{2\lambda^2}{(2\pi)^4}\Big[1-\frac{3}{2}\zeta(3)\Big(\frac{\lambda}{4\pi^2}\Big)^2+5\zeta(5)\Big(\frac{\lambda}{4\pi^2}\Big)^3 -\frac{245}{16}\zeta(7)\Big(\frac{\lambda}{4\pi^2}\Big)^4 +\frac{189}{4}\zeta(9)\Big(\frac{\lambda}{4\pi^2}\Big)^5\nonumber \\  && \hspace{1cm}  -\frac{38115}{256}\zeta(11)\Big(\frac{\lambda}{4\pi^2}\Big)^6+\frac{61347}{128}\zeta(13)\Big(\frac{\lambda}{4\pi^2}\Big)^7-\frac{6441435}{4096}\zeta(15)\Big(\frac{\lambda}{4\pi^2}\Big)^8+\cdots\Big] \nonumber \\ \\
\label{TT:44}
\langle T_4\overline{T}_4\rangle &=& \frac{4\lambda^4}{(2\pi)^8}\Big[1 -\frac{35}{32}\zeta(7)\Big(\frac{\lambda}{4\pi^2}\Big)^4 +\frac{63}{8}\zeta(9)\Big(\frac{\lambda}{4\pi^2}\Big)^5  -\frac{2541}{64}\zeta(11)\Big(\frac{\lambda}{4\pi^2}\Big)^6 +\frac{5577}{32}\zeta(13)\Big(\frac{\lambda}{4\pi^2}\Big)^7\nonumber \\  && \hspace{1cm} -\frac{2927925}{4096}\zeta(15)\Big(\frac{\lambda}{4\pi^2}\Big)^8+\cdots\Big] \nonumber \\ \\
\label{TT:66}
\langle T_6\overline{T}_6\rangle&=&\frac{6\lambda^6}{(2\pi)^{12}}\Big[1  -\frac{231}{256}\zeta(11)\Big(\frac{\lambda}{4\pi^2}\Big)^6+\frac{1287}{128}\zeta(13)\Big(\frac{\lambda}{4\pi^2}\Big)^7-\frac{289575}{4096} \zeta(15) \Big(\frac{\lambda}{4\pi^2}\Big)^8+\cdots\Big] \nonumber \\ \\
\label{TT:88}
\langle T_8\overline{T}_8\rangle&=&\frac{8\lambda^8}{(2\pi)^{16}}\Big[1-\frac{6435}{8192}\zeta(15)\Big(\frac{\lambda}{4\pi^2}\Big)^8+\cdots\Big]\nonumber \\ 
\end{eqnarray}
%
%
It is interesting to note that these correlators are compatible with a generic form

\begin{equation}
\label{TTgeneral}
\langle T_n\overline{T}_n\rangle = \frac{n\lambda^n}{(2\pi)^{2n}}\,\Big[1+\mathcal{O}(\zeta(2n-1)\lambda^n)\Big]\, ,
\end{equation}
which exhibits a remarkable cancellation of several loop Feynman diagrams, as one might naively  expect that all correlators would contain corrections starting  with $\zeta(3)\lambda^2$.

Note the analog result in the  difference in the correlator \eqref{gyb}. It must be stressed  that this result takes into account only linear terms in $\zeta(2n-1)$. It is tempting to conjecture that \eqref{TTgeneral} holds beyond linear terms in the $\zeta$'s, \textit{i.e.} that the $\langle T_n\overline{T}_m\rangle$ correlators differ from the $\mathcal{N}=4$ correlators in $O(\lambda^{2n})$, as suggested by  \eqref{TTgeneral}.
 in an expansion in $\lambda$. 
 As a check, let us compute the first non-linear term, namely the term proportional to $\zeta(3)^2$.
To leading order in the large $N$ expansion, in the $A_1$ quiver case with $\lambda_1=\lambda_2$, we find that the $\zeta(3)^2$ corrections to the 
correlators are given by

\begin{eqnarray}
&&\langle\langle \mathscr{X}_n \mathscr{X}_m\rangle\rangle =\cdots +72\, \lambda^{4+\frac{n+m}{2}}\zeta(3)^2 \, \hat{\mathscr{C}}_{2,2}\,\hat{\mathscr{C}}_{2,m}\, \hat{\mathscr{C}}_{2,n}\, ; \\ 
&&\langle\langle \mathscr{Y}_n \mathscr{Y}_m\rangle\rangle =\cdots +72\, \lambda^{4+\frac{n+m}{2}}\zeta(3)^2 \, \hat{\mathscr{C}}_{2,2}\,\hat{\mathscr{C}}_{2,m}\, \hat{\mathscr{C}}_{2,n} \, ;\\ 
&&\langle\langle \mathscr{X}_n \mathscr{Y}_m\rangle\rangle =\cdots -72\, \lambda^{4+\frac{n+m}{2}}\zeta(3)^2 \, \hat{\mathscr{C}}_{2,2}\,\hat{\mathscr{C}}_{2,m}\, \hat{\mathscr{C}}_{2,n}\, ; 
\end{eqnarray}
where $\cdots$ stands for the terms linear in $\zeta$'s computed above. 

To begin with, note that for odd correlators the $\zeta(3)^2$ correction immediately vanishes. In addition, it is clear that both towers of twisted and untwisted operators are still orthogonal, since $\langle\langle \mathscr{U}_n \mathscr{T}_m\rangle\rangle=0$. Moreover, as for the $\langle U_n\overline{U}_m\rangle$ correlator, the structure above (equal contribution, but opposite sign for the $\langle\langle \mathscr{X}_n\mathscr{X}_m\rangle\rangle$ and $\langle\langle \mathscr{X}_n\mathscr{Y}_m\rangle\rangle$ correlators) implies that there is no correction to \eqref{UUgeneral}, as expected.
Finally, let us consider the $\langle T_n\overline{T}_m\rangle$ correlator. Note first that, aside from the $\lambda^{\frac{n+m}{2}}$ overall scaling of the correlator, the $\zeta(3)^2$ correction starts with a relative $\lambda^4$ in the term inside the brackets in  \eqref{TTgeneral}. 
Thus, it can potentially contribute for $n\geq 4$. Remarkably, one can check that such correction cancels for all correlators with $n \geq 4$. 
Thus, this supports the conjecture that \eqref{TTgeneral} holds not only
for the terms which are linear in $\zeta(2n-1)$, but that
 the first correction to the $\mathcal{N}=4$ result for the $\langle T_n\overline{T}_m\rangle$ correlator is proportional to $\lambda^{2n}$ as in \eqref{TTgeneral}. While this is a conjecture (since the much harder non-linear terms in all $\zeta$'s are unknown to us), it is remarkable that, at least for the linear  terms in $\zeta(2n-1)$ and including $\zeta(3)^2$, as the dimension grows, the correlator becomes closer and closer to that in $\mathcal{N}=4$ SYM (in turn identical to the free theory).

\subsection{The general case of $A_{n-1}$ quiver gauge theories}\label{sectcorrelatorAn}

Inspired by the $A_1$ case, let us now consider the perturbative orbifold case where $\lambda_I=\lambda$ for all gauge groups. 
It is then natural to define untwisted and twisted sectors as \cite{Gukov:1998kk}

\begin{equation}
\mathscr{U}_n=\frac{1}{\sqrt{n}}\sum_K\mathscr{X}_n^K\,,\qquad \mathscr{T}^I_n=\frac{1}{\sqrt{2}}\Big( \mathscr{X}_n^I-\mathscr{X}_n^{I+1}\Big)\, .
\end{equation}
Note that, as expected, there are $n-1$ twisted sectors. Then

\begin{eqnarray}
\langle\langle \mathscr{U}_n\mathscr{T}_m^I\rangle\rangle &=& \frac{1}{\sqrt{2n}}\Big( \langle\langle \mathscr{X}_n^{I-1}\mathscr{X}_m^I\rangle\rangle+\langle\langle \mathscr{X}_n^{I}\mathscr{X}_m^I\rangle\rangle+\langle\langle \mathscr{X}_n^{I+1}\mathscr{X}_m^I\rangle\rangle \nonumber \\ && -\langle\langle \mathscr{X}_n^{I}\mathscr{X}_m^{I+1}\rangle\rangle-\langle\langle \mathscr{X}_n^{I+1}\mathscr{X}_m^{I+1}\rangle\rangle-\langle\langle \mathscr{X}_n^{I+2}\mathscr{X}_m^{I+1}\rangle\rangle\Big)\,;
\end{eqnarray}
which obviously vanishes since for $\lambda_I=\lambda$ we have that both $\langle\langle \mathscr{X}_n^{I}\mathscr{X}_m^{I+1}\rangle\rangle$ and $\langle\langle \mathscr{X}_n^{I}\mathscr{X}_m^I\rangle\rangle$ are equal for all $I$'s. Thus, we can perform the Gram-Schmidt in the $\mathscr{U}_n$ tower separately. To that mattter

\begin{equation}
\langle\langle \mathscr{U}_n\mathscr{U}_m\rangle\rangle= \frac{1}{n}\sum_{K,L} \left( \langle\langle \mathscr{X}_n^K\mathscr{X}_m^L\rangle\rangle=\frac{1}{n}\sum_K  \langle\langle \mathscr{X}_n^K\mathscr{X}_m^K\rangle\rangle+ \langle\langle \mathscr{X}_n^K\mathscr{X}_m^{K-1}\rangle\rangle+ \langle\langle \mathscr{X}_n^K\mathscr{X}_m^{K+1}\rangle\rangle \right)\, .
\end{equation}
Since at the perturbative orbifold point all couplings coincide, this is just equal to

\begin{equation}
\langle\langle \mathscr{U}_n\mathscr{U}_m\rangle\rangle=  \langle\langle \mathscr{X}_n^L\mathscr{X}_m^L\rangle\rangle+ \langle\langle \mathscr{X}_m^{L} \mathscr{X}_n^{L+1}\rangle\rangle+ \langle\langle \mathscr{X}_n^L\mathscr{X}_m^{L+1}\rangle\rangle\, ,
\end{equation}
for some fixed $L$. Using our explicit formulas we get

\begin{eqnarray}
\langle\langle \mathscr{U}_n\mathscr{U}_m\rangle\rangle&=& \langle\langle \mathscr{X}_{n} \mathscr{X}_{m}\rangle\rangle_{\rm \small SCQCD}(\lambda) +\sum_{q=2}^\infty\sum_{k=1}^{q-1}\frac{(-1)^{q}}{q}\zeta(2q-1)\binom{2q}{2k}\lambda^{m+n+q}\nonumber\\
&& \Big(\hat{\mathscr{C}}_{2k}\hat {\mathscr{C}}_{2q-2k,n,m} +\hat{\mathscr{C}}_{2q-2k}\hat{\mathscr{C}}_{2k,n,m}+\hat{\mathscr{C}}_{m,2k}\hat{\mathscr{C}}_{2q-2k,n} +\hat{\mathscr{C}}_{n,2k}\hat{\mathscr{C}}_{2q-2k,m} \Big)\, .\nonumber\\
\end{eqnarray}
Comparing with \eqref{N=2}, here we recognize in the second line $\hat{M}^{(1)}$, so that $\langle\langle \mathscr{U}_n\mathscr{U}_m\rangle\rangle=\mathscr{C}_{n,m}$, which coincides with the $\mathcal{N}=4$ correlator on $\mathbb{S}^4$. Therefore, for the $\mathbb{R}^4$ correlators, we finally  recover the $\mathcal{N}=4$ result

\begin{equation}
\langle U_nU_m\rangle=\frac{n\lambda^n}{(2\pi)^{2n}}\delta_{n,m}\,  .
\end{equation}

On the other hand, the twisted sectors are now more intricate, as they mix in a non-trivial way; that is, $\langle\langle \mathscr{T}^I_n\mathscr{T}_m^J\rangle\rangle$ does not only vanish if $I=J$. Therefore, the orthogonalization will mix the various twisted sectors.

\section{Conclusions}\label{conclusions}

In this paper we have studied several aspects of $\mathcal{N}=2$ necklace quiver gauge theories. These theories are particularly interesting given that they admit a weakly curved gravity dual in terms of the geometry $AdS_5\times S^5/\mathbb{Z}_n$.
This permits to carry out accurate tests of AdS/CFT duality with the same level of precision as in the duality between $\mathcal{N}=4$ SYM and superstring theory on  $AdS_5\times S^5$ (see {\it e.g.} \cite{Gadde:2010zi}). As one step further to  explore this duality, in this paper we have computed new exact observables in the quiver theory, namely two-point correlation functions, by using supersymmetric localization.

In the limit where all couplings but one vanish, correlation functions in  the quiver gauge theories 
reduce to those of $\mathcal{N}=2$ superconformal QCD, 
which, however, does not have a simple gravity dual where one can rely on classical supergravity calculations. Thus, our results can also be used to test
what common properties are shared between theories
without a classical gravity dual 
and  gauge theories with such a gravity dual. This can be very interesting in order to elucidate the features that the putative holographic dual to $\mathcal{N}=2$ SCQCD should exhibit.

We have found that the partition function for necklace quiver theories at infinite $N$ exhibits a rather interesting ``modular" structure allowing one to construct the partition function by adding a factor of the $\mathcal{N}=2$ SCQCD partition function for each
node 
and a factor of $\mathfrak{F}(\lambda_I, \lambda_{I+1})$ 
for each link (\textit{c.f.}  \eqref{ZAn-1}), thus extending the result in \cite{Azeyanagi:2013fla} valid for the perturbative orbifold point when all couplings are equal. We should stress that this happens in the final result, that is, after carrying out the integration over $a_i^I$ (which amounts to a full computation of the functional integral that defines the partition function).
This interesting structure seems to hold for general $\mathcal{N}=2$ theories admitting a Lagrangian description, since our derivation extends in a straightforward way to the generic
$\mathcal{N}=2$ partition functions on $\mathbb{S}^4$ computed by Pestun \cite{Pestun:2007rz}.
 It would be very interesting to study each such building block
 $\mathfrak{F}(\lambda_I, \lambda_{I+1})$ 
 in the limit where all $\lambda_I\to \infty $ with fixed $\lambda_I/\lambda_J$.
 This can be presumably studied with the techniques of \cite{Passerini:2011fe,Fraser:2015xha} and could lead to a very interesting
 test of AdS/CFT holography (generalizing our discussion of section 2.5 to the case of different couplings $\lambda_I \neq \lambda_{J}$).
 In addition,  one could investigate whether, by taking suitable limits in the $\lambda_I$'s, information about more general class $\mathcal{S}$ theories  can be found \cite{Gaiotto:2009we}.

We have initiated the study of correlation functions in necklace quiver gauge theories at large $N$. Following the method proposed in \cite{Gerchkovitz:2016gxx}, one computes the corresponding correlators in the  matrix models for the theory on $\mathbb{S}^4$ and then runs a Gram-Schmidt procedure to remove the anomalous operator mixtures (whose $AdS$ counterpart would be, \textit{per se}, very interesting to elucidate). We have computed all terms in the planar expansion that have
coefficients with linear $\zeta$ dependence. As a by-product, we have computed the analogous quantity in $\mathcal{N}=2$ SCQCD, extending previous results in the literature. This allowed us to test the extent to which correlators can be described in terms of $\mathcal{N}=4$ correlators by means of an effective coupling.
We have shown that there is no universal coupling replacement by which one can express the  two-point
correlation functions in  $\mathcal{N}=2$ SCQCD in terms of the $\mathcal{N}=4$ ones. Since the SCQCD correlation functions arise as a special limit of the
quiver correlation function (upon setting all but one coupling to zero), this result implies that
it is neither possible to express the  correlation functions of the quiver gauge theory
in terms of the $\mathcal{N}=4$ correlation functions with a universal effective
coupling. This is implied by the results in \eqref{TTgeneral}, \eqref{UUgeneral}. In particular, note that  the $SU(2,1|2)$ basis associated with scalars in each vector multiplet used in \cite{Mitev:2014yba,Fraser:2015xha,Mitev:2015oty} is just sum and difference of the $U_n, T_n$.
One could also directly choose a basis with two towers of operators associated each to a vector multiplet, \textit{i.e.} of the form $\mathscr{A}_n\sim \mathscr{X}_n+\cdots$, $\mathscr{B}_n\sim \mathscr{Y}_n+\cdots$; where the $\cdots$ stands for mixing with lower-dimensional operators within the same tower as well as with lower-dimensional operators of the other tower (that is, in general $\mathscr{X}_n$ may mix with, say, $\mathscr{Y}_{n-2}$). Of course, in the perturbative orbifold case $\mathscr{A}_n$ only contains $\mathscr{X}_n$, while $\mathscr{B}_n$ only contains $\mathscr{Y}_n$. Then, upon performing the Gram-Schmidt procedure, as in the case of $\mathcal{N}=2$ SCQCD, one finds that the resulting correlators cannot be expressed in terms of $\mathcal{N}=4$ correlation functions with a universal effective coupling. 
 
The perturbative expansion for correlators  exhibits an interesting structure.
In the case of $\mathcal{N}=2$ SCQCD, we found that the observable-dependent effective  coupling  in section \ref{SCQDrep} satisfies the
rather intriguing property   (\ref{replacementdifference}), (\ref{gyb}), checked for the first few operators up to $n=8$. 

In the $A_1$ quiver case, it turns out to be very useful to introduce untwisted and twisted sector operators. For them, at the perturbative orbifold limit $\lambda_1=\lambda_2=\lambda$, we find that their correlators either reproduce the $\mathcal{N}=4$ result (untwisted sector) or exhibit striking cancellations in planar loop Feynman diagrams yielding to the conjectured general formula \eqref{TTgeneral} (twisted sector). 
It would be very interesting to confirm these conjectures and elucidate the field theory reason for this, maybe along the lines of \cite{Bershadsky:1998mb,Bershadsky:1998cb}. 

From a holographic perspective, the result of the untwisted sector can be argued as follows.  
In the description in terms of D3 branes on $\mathbb{C}^3$, the orbifold can be taken to act in the $\mathbb{C}^2$ transverse to the $\mathbb{C}$ which is associated with the chiral field in the $\mathcal{N}=4$ theory.
Thus, in this sector,  
the $AdS_5\times S^5$ holographic calculation of \cite{Lee:1998bxa}
for CPO correlators in the $\mathcal{N}=4$ theory
should also apply to the orbifold case.

In the twisted sector, to the extent we checked, the correlators \eqref{TTgeneral}  seem to become ``closer" to the  $\mathcal{N}=4$ correlators the larger is the dimension of the operators:
 two-point correlators of operators of  dimension $n$ differ from the corresponding $\mathcal{N}=4$ correlators by   $O(\lambda^{2n} )$ in the weak coupling, planar expansion.
 This implies the remarkable cancellation of $n-1$ loop Feynman diagrams.
 It would be very interesting to clarify the nature of these cancellations. 
 
One may wonder what happens away from the perturbative orbifold case. Explicit computations with the above formulas seem to show no special cancellations.
The fact that only $\lambda_1=\lambda_2=\lambda$ shows cancellations might be connected to the failure of integrability away from $\lambda_1=\lambda_2$ (or the SCQCD limit $\lambda_2=0$) reported in \cite{Gadde:2010zi}.

In this paper we have studied the weak 't Hooft coupling regime of the necklace theories at large $N$. 
It would be very interesting to explore the strong coupling regime. This can be done by standard saddle-point techniques and might lead
to revealing new tests of holographic dualities.

\section*{Acknowledgements}

We would like to thank N. Bobev for useful discussions. J.G.R. would like to thank the Department of Physics, FCEN of Universidad de Buenos Aires, for hospitality during the course of this work.
D.R-G si partly supported by the Ramon y Cajal grant RYC-2011-07593 as well as the EU CIG grant UE-14-GT5LD2013-618459.  A. P is supported by the Asturias Government SEVERO OCHOA grant BP14-003. D.R-G and A.P acknowledge support from the Asturias Government grant FC-15-GRUPIN14-108 and Spanish Government grant MINECO-16-FPA2015-63667-P. J.G.R. acknowledges financial support from projects  FPA2013-46570,   2014-SGR-104 and  MDM-2014-0369 of ICCUB (Unidad de Excelencia `Mar\'ia de Maeztu').

\begin{appendix}

\section{Notation}\label{Notation}

In this appendix we compile a description of the notation used in the main text. We are interested on correlators of CPO's in $\mathbb{R}^4$ for necklace quiver theories. Denoting the scalar in the vector multiplet of the $I$'th gauge group as $X_{I,I}$, we shall consider single-trace chiral primary operators

\begin{equation}
X_n^I\equiv {\rm Tr}X_{I,I}^n\, .
\end{equation}
Correlation functions on $\mathbb{R}^4$ will be denoted as $\langle\cdot\rangle$. When quoting such correlators, we will omit their space dependence,
$|x_1-x_2|^{-n_1-n_2}$, which is, as usual, determined by the conformal dimensions.

In the associated  matrix model for the theory on $\mathbb{S}^4$, we can define an analog to $X^I_n$. 
However, as discussed in the main text, due to the conformal anomaly, on $\mathbb{S}^4$ the $X_n^I$ mix in a non-trivial way, which means that the $\mathbb{S}^4$ correlators, denoted by $\langle\langle X_n^I \bar X_m^L\rangle\rangle$, are not proportional to $\delta_{n,m}$. 
Recall that our 2-point functions are extremal, which means that we compute $\langle X_n\overline{X}_n\rangle$.  Note however that in $\mathbb{S}^4$ the $X_n$ avatar is real, and thus the hermitean conjugation can be dropped. 
Inside the $\mathbb{S}^4$ correlators  $\langle\langle\cdot\rangle\rangle$  the inserted operators $X_n^I$ are understood to be those of $\mathbb{S}^4$. By the same token,  inside the $\mathbb{R}^4$ correlator  $\langle \cdot\rangle$  the inserted $X_n^I$  are understood to be those of $\mathbb{R}^4$.

As discussed, in the $\mathbb{S}^4$ matrix model the $X_n^I$ mix non-trivially. In particular, they mix with the identity and thus acquire a VEV. It is
convenient to introduce VEV-less operators as

\begin{equation}
\mathscr{X}_n^I\equiv X_n^I-\langle\langle X_n^I\rangle\rangle\, .
\end{equation}
In particular, the $\mathbb{S}^4$ two-point function of the $\mathscr{X}_n^I$ is just the connected correlator of the $X_n^I$.

Upon expanding the one-loop determinants we can write the correlators (as well as the partition function) of interest in terms of quantities of the $\mathcal{N}=4$ theory, which will be denoted by a subscript $_0$. 
As in \eqref{Cs}, we introduce the $\mathcal{N}=4$ quantities

\begin{eqnarray}
\langle\langle X_n\rangle\rangle_0 = C_n(g_I)&=&\frac{1}{Z_{\mathcal{N}=4}(g_I)}\int d^{N-1} a\ e^{-\frac{8\pi^2}{g_I^2} \sum a_i^2} \Big(\sum_i a_i^n\Big)\,, 
\nonumber \\  \langle\langle X_n X_m\rangle\rangle_0=C_{m,n}(g_I)&=&\frac{1}{Z_{\mathcal{N}=4}(g_I)}\int d^{N-1}
a \ e^{-\frac{8\pi^2}{g_I^2} \sum a_i^2}\, \Big(\sum_i a_i^m\Big)\,\Big(\sum_i a_i^n\Big)\, 
\label{wer}
\end{eqnarray}
etc., and their connected counterpart \eqref{mathscrCs}

\begin{equation}
\langle\langle X_{m_1}\cdots X_{m_n}\rangle\rangle_0^c= \mathscr{C}_{m_1,\cdots,m_n}(g_I)=\frac{\partial}{\partial h_{I,m_1}}\cdots \frac{\partial}{\partial h_{I,m_n}}\mathcal{F}(g_I,\,\{h_{I,A}\})\Big|_{\{h_{I,A}=0\}}\ ,
\end{equation} 
with $\mathcal{F}=-\ln {Z}_{\mathcal{N}=4}(g_I,\,\{h_{i,A}\})$.
The latter deformed partition function for $\mathcal{N}=4$ is defined by adding sources for all single-trace chiral primary operators,

\begin{equation}
{Z}_{\mathcal{N}=4}(g_I,\,\{h_{i,A}\})=\int d^{N-1}a \ \Delta(a)\,e^{-\frac{8\pi^2}{g_I^2}\sum_ia_i^2+\sum_{A=2} h_{I,A} \sum_i  a_i^A}\ .
\end{equation}

The large $N$ limit is taken  as usual at fixed ' 't Hooft coupling $\lambda_I\equiv N g_I^2$.
A hat $\hat{}$ over $\mathscr{C}_{m_1,\cdots,m_n}$ will refer to that quantity 
evaluated at 't Hooft coupling equal to one.
On the other hand, the superscript $I$ on the    $\mathcal{N}=4$ correlators, i.e. $C_{m_1,\cdots,m_n}^I$, or the connected correlator $\mathscr{C}_{m_1,\cdots,m_n}^I$, indicate that they are computed
 in  $\mathcal{N}=4$ SYM with $\lambda_I$.

In the case of $\mathcal{N}=2$ SCQCD , since we just have one node, we can drop  superscripts  ${}^I$ and call $X_{1,1}\equiv X$.
In the case of the $A_1$ theory, given that we have only two nodes, for the sake of clarity in the formulas we define $X_{1,1}\equiv X$ and $X_{2,2}=Y$ (and similarly VEV-less operators $\mathscr{X}$, $\mathscr{Y}$).

\section{Large $N$ correlators in $\mathcal{N}=4$ SYM} \label{correlatorsN=4}

Correlators of CPO's in the large $N$ limit for ${\cal N}=4$ SYM have been computed in
 \cite{Rodriguez-Gomez:2016ijh,Rodriguez-Gomez:2016cem}. 
 We will make use of the following formulas:

\begin{eqnarray}
&&\mathscr{C}_{2r}=N\Big(\frac{\lambda}{(2\pi)^2}\Big)^r\frac{\Gamma(r+\frac{1}{2})}{\sqrt{\pi}\,\Gamma(r+2)} \ ; 
\qquad\  \mathscr{C}_{2r+1}=0\ ;\\
&&\mathscr{C}_{2n,2r}=\Big(\frac{\lambda}{(2\pi)^2}\Big)^{n+r}\frac{\Gamma(n+\frac{1}{2})\Gamma(r+\frac{1}{2})}{\pi\,(n+r)\Gamma(n)\Gamma(r)}\, 
;
\nonumber \\
&&\mathscr{C}_{2n+1,2r+1}=\Big(\frac{\lambda}{(2\pi)^2}\Big)^{n+r+1}\frac{\Gamma(n+\frac{3}{2})\Gamma(r+\frac{3}{2})}{\pi\,(n+r+1)\Gamma(n+1)\Gamma(r+1)}\, ; \\
&& \mathscr{C}_{2m,2n,2r}=\frac{1}{N}\Big(\frac{\lambda}{(2\pi)^2}\Big)^{m+n+r} \frac{\Gamma(m+\frac{1}{2})\Gamma(n+\frac{1}{2})\Gamma(r+\frac{1}{2})}{\pi^{\frac{3}{2}}\,\Gamma(m)\Gamma(n)\Gamma(r)}\, ; \\
&&\nonumber \mathscr{C}_{2m,2n,2r,2s}=\frac{1}{N^2}\Big(\frac{\lambda}{(2\pi)^2}\Big)^{m+n+r+s} \frac{\Gamma(m+\frac{1}{2})\Gamma(n+\frac{1}{2})\Gamma(r+\frac{1}{2})\Gamma(s+\frac{1}{2})\,(n+m+r+s-1)}{\pi^2\,\Gamma(m)\Gamma(n)\Gamma(r)\Gamma(s)}\, .\\
\end{eqnarray}
Note in particular that

\begin{equation}
\mathscr{C}_{m_1,\cdots,m_n}\sim N^{2-n}\, .
\end{equation}

\section{The partition function for the $A_1$ theory}\label{appendixZA1}

In this appendix we give further details on the computation of the partition Starting with (\ref{eef})  and using (\ref{expansionH}), we find
function for the $A_1$ case. It is convenient to rename $a_i^1\rightarrow a_i$
and $a_i^2\rightarrow b_i$ in \eqref{pathintegralZAn-1}.
Starting with (\ref{eef})  and using (\ref{expansionH}), we find

\begin{equation}
 f=-\sum_{i,j}\sum_{n=2}^\infty \frac{(-1)^n}{n}\zeta(2n-1) \Big((a_i-a_j)^{2n}+(b_i-b_j)^{2n}-2(a_i-b_j)^{2n}\Big)\, .
\end{equation}
i.e.

\begin{equation}
 f=-\sum_{n=2} ^\infty \frac{(-1)^n}{n}\zeta(2n-1) \sum_{k=0}^{2n} (-1)^k {{2n}\choose{k}}\sum_i(a_i^{2n-k}-b_i^{2n-k})\sum_j(a_j^{k}-b_j^{k})\, .
\end{equation}
Recall now that $a_i$, $b_i$ stand respectively for the eigenvalues of the $X_{1,1}$ and $X_{2,2}$ adjoint scalars in the quiver. In order to ease notation, denoting by $X$ the adjoint scalar of the first group and by $Y$ the adjoint scalar of the second, this is

\begin{equation}
 f=-\sum_{n=2}^\infty \sum_{k=0}^{2n} \frac{(-1)^{n+k}}{n}\zeta(2n-1) {{2n}\choose{k}} ({\rm Tr}X^{2n-k}-{\rm Tr}Y^{2n-k})({\rm Tr}X^{k}-{\rm Tr}Y^{k})
\end{equation}
Recall now that $Z_{\rm 1-loop}=e^f$, which must be inserted in the integral \eqref{ZA1}. Expanding the exponential $e^f=1+f+\cdots$, when inserted in \eqref{ZA1}, the ``1" will give the product of two copies of the partition function of the $\mathcal{N}=4$ theory with the appropriate coupling, that is, $Z_{\mathcal{N}=4}(\lambda_1)Z_{\mathcal{N}=4}(\lambda _2)$. Likewise, the linear term in $f$ involves the integrals of

\begin{equation}
 ({\rm Tr}X^{2n-k}-{\rm Tr}Y^{2n-k})({\rm Tr}X^{k}-{\rm Tr}Y^{k})
 \end{equation}
In terms of the $C_{m_1\cdots m_n}$ defined in \eqref{wer}, the integrated linear term with $f$ reads

\begin{equation}
Z_{\mathcal{N}=4}(\lambda_1)Z_{\mathcal{N}=4}(\lambda_2) \Big( C_{2n-k,k}(\lambda_1)-C_{2n-k}(\lambda_1) C_k(\lambda_2)-C_{2n-k}(\lambda_2)C_k(\lambda_1)+C_{2n-k,k}(\lambda_2)\Big)
 \end{equation}
The $C$'s are easily re-written in terms of the $\mathscr{C}$'s in appendix \ref{correlatorsN=4}, so that one can check that the integral of the linear term in $f$ reads (we omit the overall factor $Z_{\mathcal{N}=4}(\lambda_1)Z_{\mathcal{N}=4}(\lambda_2)$)

\begin{equation}
\mathscr{C}_{2n-k,k}(\lambda_1)+\mathscr{C}_{2n-k}(\lambda_1)\mathscr{C}_k(\lambda_1)-\mathscr{C}_{2n-k}(\lambda_1) \mathscr{C}_k(\lambda_2)-\mathscr{C}_{2n-k}(\lambda_2)\mathscr{C}_k(\lambda_1)+\mathscr{C}_{2n-k,k}(\lambda_2)+\mathscr{C}_{2n-k}(\lambda_2)\mathscr{C}_k(\lambda_2)
 \end{equation}
The $\mathscr{C}$'s are given in appendix \ref{correlatorsN=4} in terms of the 't Hooft coupling. Note in particular that $\mathscr{C}_{m,n}$ is subleading in $N$ with respect to $\mathscr{C}_m$. Thus, the leading term above is

\begin{eqnarray}
\mathscr{C}_{2n-k}(\lambda_1)\mathscr{C}_k(\lambda_1)-\mathscr{C}_{2n-k}(\lambda_1) \mathscr{C}_k(\lambda_2)-\mathscr{C}_{2n-k}(\lambda_2)\mathscr{C}_k(\lambda_1)+\mathscr{C}_{2n-k}(\lambda_2)\mathscr{C}_k(\lambda_2)&&\\ \nonumber =[\mathscr{C}_{2n-k}(\lambda_1)-\mathscr{C}_{2n-k}(\lambda_2)][\mathscr{C}_k(\lambda_1)- \mathscr{C}_k(\lambda_2)]\, .&&
 \end{eqnarray}
This can be extended to all powers of $f$, i.e. in the large $N$ limit, and the dominant term in $f^n$ is obtained by replacing
the operators  ${\rm Tr}X^{2n}$, ${\rm Tr}Y^{2m}$ by their VEV's, $\mathscr{C}_{2n}(\lambda_1)$, $\mathscr{C}_{2m}(\lambda_2)$, leading to $\ln f\sim N^2$.   This gives (\ref{partiZZ}), (\ref{fZZ}).

\section{Two-point correlators in ${\cal N}=2$ SCQCD }\label{correlatorsN=2}

In this appendix we offer the details of the computation of correlators in $\mathcal{N}=2$ superconformal QCD including all terms linear in $\zeta(2n-1)$ as well as the first non-linear term in the $\zeta(2n-1)$'s, namely the term with coefficient $\zeta(3)^2$. In the case of $\mathcal{N}=2$ superconformal QCD, the one-loop contribution to the matrix model is

\begin{equation}
Z_{\rm 1-loop} = 
e^{\sum_{i,j=1}^N \left(\ln H(a_i-a_j)-\ln H(a_i) - \ln H(a_j)\right)}\, .
\end{equation}
Using \eqref{expansionH}, we can write the exponent as

\begin{equation}
\sum_{i,j} \left(\ln H(a_i-a_j)-\ln H(a_i) - \ln H(a_j)\right)= \sum_{i,j=1}^N \sum_{q=2}^\infty (-1)^q\frac{\zeta(2q-1)}{q} [a_i^{2q}+a_j^{2q} - (a_i-a_j)^{2q}]\, .
\end{equation}
Expanding the binomials, and using that, in $SU(N)$, $\sum_i a_i=0$, we obtain

\begin{equation}
Z_{\rm 1-loop}=e^{-\sum_{i,j}\sum_{q=2}^\infty\sum_{k=2}^{2q-2} (-1)^{q+k}\frac{\zeta(2q-1)}{q} \binom{2q}{k} a_i^k a_j^{2q-k}}\, .
\end{equation}
The terms which are linear in $\zeta(2q-1)$ are given by

\begin{equation}
Z_{\rm 1-loop} = 1 -\sum_{i,j=1}^N\sum_{q=2}^\infty \sum_{k=2}^{2q-2}(-1)^{q+k}\frac{\zeta(2q-1)}{q} \binom{2q}{k} a_i^k a_j^{2q-k}+...\, .
\end{equation}
where dots stand for the rest of the terms which are product of  $\zeta$'s.
Thus, we get

\begin{equation}
Z=Z_0\ \Big(1-\sum_{q=2}^\infty \sum_{k=2}^{2q-2} (-1)^{q+k}\frac{\zeta(2q-1)}{q} \binom{2q}{k}C_{k,2q-k}+...\Big)\, .
\end{equation}
Hence, keeping only the linear terms in $\zeta $'s, , we have

\begin{equation}
\langle\langle X_nX_m\rangle\rangle = C_{n,m} -\sum_{q=2}^\infty \sum_{k=2}^{2q-2} (-1)^{q+k}\frac{\zeta(2q-1)}{q} \binom{2q}{k}\Big( C_{k,2q-k,n,m}-C_{k,2q-k}C_{n,m}\Big)\,.
\end{equation}

\noindent Likewise

\begin{equation}
\langle\langle X_n\rangle\rangle = C_n -\sum_{q=2}^\infty \sum_{k=2}^{2q-2} (-1)^{q+k}\frac{\zeta(2q-1)}{q} \binom{2q}{k}\Big( C_{k,2q-k,n}-C_{k,2q-k} C_n \Big)\, .
\end{equation}

Combining these two formulas, we obtain, for the  connected correlator, 

\begin{equation}
\langle\langle \mathscr{X}_n\mathscr{X}_m\rangle\rangle=\mathscr{C}_{n,m} -\sum_{q=2}^\infty \sum_{k=2}^{2q-2} (-1)^{q+k}\frac{\zeta(2q-1)}{q} \binom{2q}{k} M^{(1)}_{k,q,n,m}\ ,
\end{equation}
where

\begin{equation}
M^{(1)}_{k,q,n,m} = C_{k,2q-k,n,m}-C_{k,2q-k}  C_{n,m}   -  C_{k,2q-k,n} C_m-C_{k,2q-k,m}C_n+2C_{k,2q-k} C_n C_m\,.
\end{equation}
In terms of connected correlators this reads

\begin{equation}
M^{(1)}_{k,q,n,m}=\mathscr{C}_{k,2q-k,n,m} + \mathscr{C}_{k,n,m}\mathscr{C}_{2q-k} + \mathscr{C}_{2q-k,n,m}\mathscr{C}_{k}  +\mathscr{C}_{k,n} \, \mathscr{C}_{2q-k,m} +\mathscr{C}_{2q-k,n} \mathscr{C}_{k,m}\ .
\end{equation}
Recalling that $\mathscr{C}_{k,m}=O(1)$ and $\mathscr{C}_{k}=O(N)$, to leading order in $N$ we finally obtain

\begin{equation}
\label{emeuno}
M^{(1)}_{k,q,n,m}= \mathscr{C}_{k,n,m}\mathscr{C}_{2q-k} + \mathscr{C}_{2q-k,n,m}\mathscr{C}_{k}  +\mathscr{C}_{k,n} \mathscr{C}_{2q-k,m} +\mathscr{C}_{2q-k,n} \mathscr{C}_{k,m}\, .
\end{equation}

We can easily extend the above computation to include the first non-linear term in the $\zeta$'s, namely the one proportional to $\zeta(3)^2$. Note that in the expansion of the one-loop factor we have

\begin{equation}
Z_{\rm 1-loop} = 1 -\sum_{i,j=1}^N\sum_{q=2}^\infty\sum_{k=2}^{2q-2} (-1)^{q+k}\frac{\zeta(2q-1)}{q} \binom{2q}{k} a_i^k a_j^{2q-k}+\frac{9}{2} \zeta(3)^2 (\sum_{i=1}^N a_i^2)^4 +\cdots\, 
\end{equation}
Therefore (we omit dots ``$\cdots$" in what follows)

\begin{equation}
Z=Z_0\ \Big(1-\sum_{q=2}^\infty \sum_{k=2}^{2q-2} (-1)^{q+k}\frac{\zeta(2q-1)}{q} \binom{2q}{k}C_{k,2q-k}+\frac{9}{2}\zeta(3)^2 C_{2,2,2,2}\Big)\, .
\end{equation}
In addition,

\begin{eqnarray}
\langle\langle X_nX_m\rangle\rangle = C_{n,m} & -&\sum_{q=2}^\infty\sum_{k=2}^{2q-2} (-1)^{q+k}\frac{\zeta(2q-1)}{q} \binom{2q}{k}\Big( C_{k,2q-k,n,m}-C_{k,2q-k}C_{n,m}\Big) \nonumber\\ & +& \frac{9}{2}\zeta(3)^2 (C_{2,2,2,2,n,m}-C_{2,2,2,2}C_{n,m}+2 C_{2,2} C_{2,2}C_{n,m}-2C_{2,2}C_{2,2,n,m})\Big) \, , \nonumber \\
\end{eqnarray}
and

\begin{eqnarray}
\langle\langle X_n\rangle\rangle = C_n &-&\sum_{q=2}^\infty \sum_{k=2}^{2q-2} (-1)^{q+k}\frac{\zeta(2q-1)}{q} \binom{2q}{k}\Big( C_{k,2q-k,n}-C_{k,2q-k} C_n \Big) \nonumber \\ & +&\frac{9}{2}\zeta(3)^2\Big(C_{2,2,2,2,n} +2C_{2,2} C_{2,2} C_n  -C_{2,2,2,2}C_n-2C_{2,2}  C_{2,2,n}\Big) \, .
\end{eqnarray}
Combining these expressions, we  obtain the following formula for the connected correlator including the $\zeta(3)^2$ order:

\begin{equation}
\langle\langle \mathscr{X}_n\mathscr{X}_m\rangle\rangle=\mathscr{C}_{n,m} -\sum_{q=2}^\infty\sum_{k=2}^{2q-2} (-1)^{q+k}\frac{\zeta(2q-1)}{q} \binom{2q}{k} M^{(1)}_{k,q,n,m}+\frac{9}{2}\zeta(3)^2\,M_{k,q,n,m}^{(2)}\, ,
\end{equation}
where

\begin{eqnarray}
\nonumber M_{k,q,n,m}^{(2)}&=& C_{2,2,2,2,n,m} - 2 C_{2,2,m} C_{2,2,n}  -2C_{2,2} C_{2,2,n,m}  -C_{2,2,2,2,n} C_m -C_{2,2,2,2,m} C_n \\ \nonumber && + 4C_{2,2} C_{2,2,n}C_m  +4 C_{2,2}C_{2,2,m}C_n   - 6 C_{2,2}C_{2,2} C_n C_m +2C_{2,2,2,2} C_m C_n  \\  && +2C_{2,2} C_{2,2} C_{n,m} -C_{2,2,2,2}C_{n,m}\, .
\end{eqnarray}

In terms of connected correlators we have 

\begin{eqnarray}
\nonumber M_{k,q,n,m}^{(2)}&=&\mathscr{C}_{2,2,2,2,n,m}+4\mathscr{C}_2\mathscr{C}_{2,2,2,n,m}+4\mathscr{C}_{2,2,n}\mathscr{C}_{2,2,m}+4\mathscr{C}_2\mathscr{C}_2 \mathscr{C}_{2,2,n,m} +4 \mathscr{C}_{2,2}  \mathscr{C}_{2,2,n,m} \\ \nonumber && +4 \mathscr{C}_{2,2,2,n} \mathscr{C}_{2,m} + 4 \mathscr{C}_{2,2,2,m}\mathscr{C}_{2,n} + 4\mathscr{C}_{2,2,2} \mathscr{C}_{2,n,m}  + 8 \mathscr{C}_2  \mathscr{C}_{2,2,n} \mathscr{C}_{2,m} + 8 \mathscr{C}_2 \mathscr{C}_{2,2,m} \mathscr{C}_{2,n} \\  && + 8 \mathscr{C}_{2,2} \mathscr{C}_{2,m} \mathscr{C}_{2,n} +8\mathscr{C}_2 \mathscr{C}_{2,2} \mathscr{C}_{2,n,m}\, .
\end{eqnarray}
Thus, the leading  term in the $1/N$ expansion in $M^{(2)}$ is given by

\begin{equation}
 M_{k,q,n,m}^{(2)}=4\mathscr{C}_2\mathscr{C}_2 \mathscr{C}_{2,2,n,m}+ 8 \mathscr{C}_2  \mathscr{C}_{2,2,n} \mathscr{C}_{2,m} + 8 \mathscr{C}_2 \mathscr{C}_{2,2,m} \mathscr{C}_{2,n}+ 8 \mathscr{C}_{2,2} \mathscr{C}_{2,m} \mathscr{C}_{2,n} +8\mathscr{C}_2 \mathscr{C}_{2,2} \mathscr{C}_{2,n,m}  \, .
\end{equation}

\section{Two-point correlators in quiver gauge theories}\label{correlatorsquiver}

In this appendix we describe the computation of correlators in quiver gauge theories including all terms linear in the $\zeta$'s. 

\subsection{General quiver}

Expanding
the one-loop factor  \eqref{generaln} in powers of $X$ and keeping only the  linear term in the $\zeta$'s , we obtain

\begin{equation}
Z_{\rm 1-loop}=1-\sum_{I=1}^n \sum_{q=2}^\infty\sum_{k=0}^{2q}\frac{(-1)^{q+k}}{q}\zeta(2q-1)\binom{2q}{k}\Big( {\rm Tr}X_{I,I}^{2q-k}{\rm Tr}X_{I,I}^k-{\rm Tr}X_{I,I}^{2q-k}{\rm Tr}X_{I+1,I+1}^k\Big)\, .
\end{equation}
From here, it follows immediately

\begin{equation}
Z=1-\sum_{I=1}^n\sum_{q=2}^\infty\sum_{k=0}^{2q}\frac{(-1)^{q+k}}{q}\zeta(2q-1)\binom{2q}{k}\Big(C^I_{2q-k,k}-C^I_{2q-k} C^{I+1}_k\Big)\, .
\end{equation}
where a superscript $I$ will refer to the corresponding quantity computed in a copy of the $\mathbb{S}^4$ matrix model with coupling $\lambda_I$. In the following we will use a simplified notation in which $X^I_k\equiv {\rm Tr} X_{I,I}^k$. In this notation, for instance

\begin{equation}
Z_{\rm 1-loop}=1-\sum_{I=1}^n
\sum_{q=2}^\infty\sum_{k=0}^{2q}\frac{(-1)^{q+k}}{q}\zeta(2q-1)\binom{2q}{k}\Big(X^I_{2q-k}X^I_k-X^I_{2q-k}X^{I+1}_k\Big)\, .
\end{equation}

Let us start by considering the connected correlator $\langle\langle \mathscr{X}^L_n \mathscr{X}^L_m\rangle\rangle$. After a straightforward but rather tedious computation one finds

\begin{equation}
\langle\langle X_n^LX_m^L\rangle\rangle = C^L_{n,m}-\sum_{q=2}^\infty \sum_{k=0}^{2q}\frac{(-1)^{q+k}}{q}\zeta(2q-1)\binom{2q}{k}
\left( A^L_{q,k,n,m} - \sum_{I\ne L, L-1}B^{L,I}_{q,k,n,m}\right)\, ;
\end{equation}
where

\begin{equation}
A^L_{q,k,n,m}=(C_{2q-k,k,n,m}^L-C^{L}_{2q-k,k}C^L_{n,m}-C^{L+1}_kC^L_{2q-k,n,m}-C^{L-1}_{2q-k}C^{L}_{k,n,m}+C^L_{2q-k} C^{L+1}_kC^L_{n,m}+C^{L-1}_{2q-k} C^{L}_kC^L_{n,m})\, ,
\end{equation}
and 

\begin{equation}
B^{L,I}_{q,k,n,m}=(C^I_{2q-k}C^{I+1}_kC^L_{n,m}-C^I_{2q-k,k}C^L_{n,m}+C^I_{2q-k,k}C^L_{n,m}-C^I_{2q-k} C^{I+1}_kC^L_{n,m}\Big)
\end{equation}\, .
Likewise

\begin{equation}
\langle\langle X_n^L\rangle\rangle = C^L_{n}-\sum_{q=2}^\infty \sum_{k=0}^{2q}\frac{(-1)^{q+k}}{q}\zeta(2q-1)\binom{2q}{k}
\left(E^L_{q,k,n} -\sum_{I\ne L, L-1} F^{L,I}_{q,k,n}\right)\, ;
\end{equation}
where now

\begin{equation}
E^L_{q,k,n}=(C_{2q-k,k,n}^L-C^{L}_{2q-k,k}C^L_{n}-C^{L+1}_kC^L_{2q-k,n}-C^{L-1}_{2q-k}C^{L}_{k,n}+C^L_{2q-k} C^{L+1}_kC^L_{n}+C^{L-1}_{2q-k} C^{L}_kC^L_{n})\, ,
\end{equation}
and

\begin{equation}
F^{L,I}_{q,k,n}=(C^I_{2q-k}C^{I+1}_kC^L_{n}-C^I_{2q-k,k}C^L_{n}+C^I_{2q-k,k}C^L_{n}-C^I_{2q-k} C^{I+1}_kC^L_{n}\Big)\, .
\end{equation}
Combining these expressions and writing the result in terms of connected correlators, we obtain

\begin{eqnarray}
\langle\langle \mathscr{X}^L_n \mathscr{X}^L_m\rangle\rangle &=& \mathscr{C}^L_{n,m}-\sum_{q=2}^\infty\sum_{j=0}^{2q}\frac{(-1)^{q+j}}{q}\zeta(2q-1)\binom{2q}{j}M^{(1)}_{q,j,n,m} \\ \nonumber && +\sum_{q=2}^\infty\sum_{k=0}^{2q}\frac{(-1)^{q+k}}{q}\zeta(2q-1)\binom{2q}{k}\Big(\mathscr{C}^{L+1}_k\mathscr{C}^L_{2q-k,n,m} +\mathscr{C}^{L-1}_{2q-k}\mathscr{C}^L_{k,n,m}\Big)\, ,
\end{eqnarray}
where $M^{(1)}_{q,k,n,m}$ is given in \eqref{emeuno}.
The first line is nothing but the $\mathcal{N}=2$ SCQCD  correlator with coupling $\lambda_L$. In the second line, only even $k$ contributes, since
$\mathscr{C}^L_{2n+1}=0$.
 Thus  we find

\begin{eqnarray}
\langle\langle \mathscr{X}^L_n \mathscr{X}^L_m\rangle\rangle 
&=& \langle\langle \mathscr{X}^L_n \mathscr{X}^L_m\rangle\rangle_{\rm \small SCQCD}(\lambda_L) 
\nonumber\\
&+&\sum_{q=2}^\infty\sum_{k=1}^{q-1}\frac{(-1)^{q}}{q}\zeta(2q-1)\binom{2q}{2k}\Big(\mathscr{C}^{L+1}_{2k}\mathscr{C}^L_{2q-2k,n,m} +\mathscr{C}^{L-1}_{2q-2k}\mathscr{C}^L_{2k,n,m}\Big)\, .
\end{eqnarray}

Now consider $\langle\langle \mathscr{X}^L_n \mathscr{X}^{L+1}_m\rangle\rangle$. A similar computation yields to

\begin{equation}
\langle\langle X_n^LX_m^{L+1}\rangle\rangle = \nonumber C_n^LC_m^{L+1}-\sum_{q=2}^\infty\sum_{k=0}^{2q}\frac{(-1)^{q+k}}{n}\zeta(2n-1)\binom{2q}{k} \  G^L_{q,k,n,m}
\end{equation}
with

\begin{eqnarray}
 \nonumber G^L_{q,k,n,m}&=&C^L_{2q-k,k,n}C_m^{L+1}+C^{L+1}_{2q-k,k,m}C_n^L- C^L_{2q-k,n}C^{L+1}_{k,m} - C^{L+1}_{2q-k,m} C^{L+2}_kC_n^L- C^{L}_{k,n} C_m^{L+1}C^{L-1}_{2q-k}\\ \nonumber && +C^{L-1}_{2n-k,k} C_n^LC_m^{L+1}- C_n^L C_m^{L+1}\Big(C^L_{2q-k,k}-C^L_{2q-k} C^{L+1}_k+C^{L-1}_{2q-k,k}-C^{L-1}_{2q-k} C^{L}_k+C^{L+1}_{2q-k,k}\\ \nonumber &&  -C^{L+1}_{2q-k} C^{L+2}_k\Big)\, ,
\end{eqnarray}
and

\begin{equation}
\langle\langle X_n^L\rangle\rangle =  C_n^L-\sum_{q=2}^\infty\sum_{k=0}^{2q}\frac{(-1)^{q+k}}{q}\zeta(2q-1)\binom{2q}{k}H^L_{q,k,n} 
\, ,
\end{equation}
with

\begin{eqnarray}
\nonumber H^L_{q,k,n}&=& C^L_{2q-k,k,n}+C^{L+1}_{2q-k,k}C_n^L -C_{k}^{L+1} C^L_{2q-k,n} -C^L_{2q-k,k}C_n^L+C^L_{2q-k} C^{L+1}_kC_n^L-C^{L+1}_{2q-k,k}C_n^L \\ \nonumber &&+C^{L-1}_{2q-k,k}C^{L}_{n} -C^L_{k,n}C^{L-1}_{2q-k} -C^{L-1}_{2q-k,k}C_n^L+C^{L-1}_{2q-k} C^{L}_kC_n^L+C^{L+1}_{2q-k} C^{L+2}_kC_n^L -C^{L+1}_{2q-k}C^{L+2}_kC_n^L\, ,\\ 
\end{eqnarray}
Using the above formulas, we finally find

\begin{eqnarray}
\langle\langle \mathscr{X}^L_n \mathscr{X}^{L+1}_m\rangle\rangle &=& \sum_{q=2}^\infty \sum_{k=0}^{2q}\frac{(-1)^{q+k}}{q}\zeta(2q-1)\binom{2q}{k}\mathscr{C}^{L+1}_{m,k}\mathscr{C}^L_{2q-k,n}\, .
\end{eqnarray}
Note that the case when one of the integers in $\hat{\mathscr{C}}_{n,m}$ vanishes is excluded. Thus, we could really restrict the sum in $k$ from 1 to $2q-1$.

It remains to consider the correlator $\langle\langle \mathscr{X}^L_n \mathscr{X}^M_m\rangle\rangle$ with $|L-M|>1$. A straightforward but long computation gives

\begin{equation}
\langle\langle \mathscr{X}^L_n\mathscr{X}^M_m\rangle\rangle = \sum_{q=2}^\infty
\sum_{k=0}^{2q}\frac{(-1)^{q+k}}{q}\zeta(2q-1)\binom{2q}{k} \Big[ J^{L,M}_{q,k,n,m} - (K^{L,M}_{q,k,n}C^M_m+K^{M,L}_{q,k,m}C^L_n)\Big]\, ,
\end{equation}
where
\begin{eqnarray}
\nonumber J^{L,M}_{q,k,n,m}&=&-C^L_{2q-k,k,n} C_m^{M}-C^M_{2q-k,k,m}C_n^L-C^{L-1}_{2q-k,k}C_n^LC_m^{M}-C^{M-1}_{2q-k,k} C_n^LC_m^{M} +C^L_{2q-k,n}C^{L+1}_kC_m^{M}\\ \nonumber && +C^{L}_{k,n} C_m^{M}C^{L-1}_{2q-k}  +C^M_{2q-k,m}C^{M+1}_kC_n^L+C^{M}_{k,m}C_n^LC^{M-1}_{2q-k} +C^L_n C^M_m\Big(C^L_{2q-k,k}-C^L_{2q-k} C^{L+1}_k\\ \nonumber && +C^{L-1}_{2q-k,k}-C^{L-1}_{2q-k} C^{L}_k+ C^M_{2q-k,k}-C^M_{2q-k} C^{M+1}_k+C^{M-1}_{2q-k,k}-C^{M-1}_{2q-k} C^{M}_k\Big)\, , \\
\end{eqnarray}
and

\begin{eqnarray}
\nonumber K^{L,M}_{q,k,n}&=&-C^L_{2q-k,k,n} -C^M_{2q-k,k}C_n^L-C^{L-1}_{2q-k,k}C_n^L-C^{M-1}_{2q-k,k} C_n^L +C^L_{2q-k,n}C^{L+1}_k+C^{L}_{k,n} C^{L-1}_{2q-k} \\ \nonumber && +C^M_{2q-k}C^{M+1}_kC_n^L+C^{M}_{k}C_n^LC^{M-1}_{2q-k} +C^L_n \Big(C^L_{2q-k,k}-C^L_{2q-k} C^{L+1}_k+C^{L-1}_{2q-k,k} -C^{L-1}_{2q-k} C^{L}_k\\ \nonumber && +C^M_{2q-k,k}-C^M_{2q-k} C^{M+1}_k+C^{M-1}_{2q-k,k}-C^{M-1}_{2q-k} C^{M}_k\Big)\, .\\
\end{eqnarray}
Thus, we see that this exactly vanishes, and hence

\begin{equation}
\langle\langle \mathscr{X}^L_n \mathscr{X}^M_m\rangle\rangle = 0\, .
\end{equation}

\subsection{The $A_1$ quiver}

Note that the $A_1$ case is slightly special, as the two nearest neighbours to one node are the same: the other node. In this case the computation of the mixed correlator must be done {\it ab initio}. We find

\begin{equation}
\langle\langle X_nY_m\rangle\rangle = C^X_nC^Y_m-\sum_{q=2}^\infty \sum_{k=0}^{2q}\frac{(-1)^{q+k}}{q}\zeta(2q-1)\binom{2q}{k}A_{q,k,n,m}^{X,Y}\, ,
\end{equation}

with

\begin{eqnarray}
 \nonumber A_{q,k,n,m}^{X,Y}&=&C^X_{2q-k,k,n}C^Y_m+C^Y_{2q-k,k,m} C^X_n-C^X_{2q-k,n} C^Y_{k,m}-C^Y_{2q-k,m}C^X_{k,n} -C^X_nC^Y_m\Big(C^X_{2q-k,k}+C^Y_{2q-k,k} \\ \nonumber && -C^X_{2q-k} C^{Y}_k-C^Y_{2q-k} C^{X}_k\Big)\, .\\
\end{eqnarray}

Likewise, we obtain for $\langle\langle X_n\rangle\rangle$,

\begin{equation}
\langle\langle X_n\rangle\rangle=C^X_n-\sum_{q=2}^\infty \sum_{k=0}^{2q}\frac{(-1)^{q+k}}{q}\zeta(2q-1)\binom{2q}{k}E_{q,k,n}^{X,Y}\, ,
\end{equation}
with

\begin{eqnarray}
 \nonumber E_{q,k,n}^{X,Y}&=&C^X_{2q-k,k,n}+C^Y_{2q-k,k} C^X_n-C^X_{2q-k,n} C^Y_{k}-C^Y_{2q-k}C^X_{k,n}  -C^X_n\Big(C^X_{2q-k,k}+C^Y_{2q-k,k}\\ \nonumber && -C^X_{2q-k} C^{Y}_k-C^Y_{2q-k} C^{X}_k\Big)\, .\\
\end{eqnarray}
and similarly for $\langle\langle Y_m\rangle\rangle$. Combining these equations, one finally finds

\begin{equation}
\langle\langle \mathscr{X}_n\mathscr{Y}_m\rangle\rangle = -\sum_{q=2}^\infty \sum_{k=0}^{2q}\frac{(-1)^{q+k}}{q}\zeta(2q-1)\binom{2q}{k} [ A_{q,k,n,m}^{X,Y} - (E_{q,k,n}^{X,Y}C^Y_m+E_{q,k,m}^{Y,X}C^X_n)]\, ;
\end{equation}
which gives

\begin{equation}
\label{A1XY}
\langle\langle \mathscr{X}_n\mathscr{Y}_m\rangle\rangle = \sum_{q=2}\sum_{k=0}^{2q}\frac{(-1)^{q+k}}{q}\zeta(2q-1)\binom{2q}{k} \Big(\mathscr{C}^X_{2q-k,n}\mathscr{C}^Y_{k,m} +\mathscr{C}^Y_{2q-k,m}\mathscr{C}^X_{k,n} \Big)\, .
\end{equation}

\end{appendix}

\end{document}